\documentclass[11pt,letter]{article}
\usepackage{fancyhdr}
\usepackage{lastpage}
\usepackage[toc,page]{appendix}
\usepackage[ruled,vlined]{algorithm2e}
\usepackage{rotating}

\SetKwInput{KwInput}{Input}                
\SetKwInput{KwOutput}{Output}              


\usepackage[parfill]{parskip} 
\usepackage[margin=1in]{geometry}
\usepackage[doublespacing]{setspace}
\usepackage[sort,round]{natbib}
\usepackage{url}
\usepackage{amsfonts,amsmath,bbm,bm,amssymb,mathtools}
\usepackage{xcolor}
\usepackage{ulem}
\usepackage{amsmath}
\usepackage[colorinlistoftodos]{todonotes}
\usepackage{caption} 
\usepackage{multirow}
\usepackage[flushleft]{threeparttable}

\usepackage[pagebackref=true,bookmarks]{hyperref}
\hypersetup{
	unicode=false,
	pdftoolbar=true,
	pdfmenubar=true,
	pdffitwindow=false,     
	pdfstartview={FitH},    
	pdftitle={Penalized GLMM},    
	pdfauthor={JSP},     
	pdfsubject={Subject},   
	pdfcreator={JSP},   
	pdfproducer={JSP}, 
	pdfkeywords={}, 
	pdfnewwindow=true,      
	colorlinks=true,       
	linkcolor=red,          
	citecolor=blue,        
	filecolor=black,      
	urlcolor=blue           
}

\usepackage{xr}
\makeatletter
\newcommand*{\addFileDependency}[1]{
  \typeout{(#1)}
  \@addtofilelist{#1}
  \IfFileExists{#1}{}{\typeout{No file #1.}}
}
\makeatother

\newcommand*{\myexternaldocument}[1]{
    \externaldocument{#1}
    \addFileDependency{#1.tex}
    \addFileDependency{#1.aux}
}

\myexternaldocument{supp}

\def\P{\text{P}}

\def\Var{\text{Var}}

\def\E{\mathbb{E}}

\def\({\left(}
\def\){\right)}
\def\[{\left[}
\def\]{\right]}

\title{Penalized generalized linear mixed models for longitudinal outcomes in genetic association studies}

\author{JULIEN ST-PIERRE\thanks{To whom correspondence should be addressed.}\\[4pt]
\textit{Department of Epidemiology, Biostatistics and Occupational Health,} \\
\textit{McGill University, Montreal, Canada}
\\[2pt]
{julien.st-pierre@mail.mcgill.ca}\\[4pt]
SAHIR RAI BHATNAGAR\\[4pt]
\textit{Department of Epidemiology, Biostatistics and Occupational Health,} \\
\textit{McGill University, Montreal, Canada}\\[4pt]
MASSIMILIANO ORRI \\[4pt]
\textit{Department of Epidemiology, Biostatistics and Occupational Health} \\
\textit{McGill University, Montreal, Canada} \\
\textit{McGill Group for Suicide Studies, Douglas Mental Health University Institute,} \\
\textit{Department of Psychiatry,}
\textit{McGill University, Montreal, Canada}\\[4pt]
MICHEL BOIVIN \\[4pt]
\textit{Research Unit On Children’s Psychosocial Maladjustment, University of Montreal, Montreal, Canada,} \\
\textit{Department of Psychology, Laval University, Quebec, Canada} \\[4pt]
JOSÉE DUPUIS \\[4pt]
\textit{Department of Epidemiology, Biostatistics and Occupational Health,} \\
\textit{McGill University, Montreal, Canada}\\
KARIM OUALKACHA\\[4pt]
\textit{Département de Mathématiques,} \\
\textit{Université du Québec à Montréal, Montreal, Canada}\\[4pt]} 

\begin{document}
\pagestyle{fancy}
\maketitle

\begin{abstract}
This work is motivated by analyses of longitudinal data collected from participants in the Quebec Longitudinal Study of Child Development (QLSCD) and the Quebec Newborn Twin Study (QNTS) to identify important genetic predictors for emotional and behavioral difficulties in childhood and adolescence. We propose a lasso penalized mixed model for continuous and binary longitudinal traits that allows the inclusion of multiple random effects to account for random individual effects not attributable to the genetic similarity between individuals. Through simulation studies, we show that replacing the estimated genetic relatedness matrix (GRM) by a sparse matrix introduces bias in the variance components estimates, but that the obtained computational gain is major while the impact on the performance of the penalized model to retrieve important predictors is negligible. We compare the performance of the proposed penalized mixed model to a standard lasso and to a univariate mixed model association test and show that the proposed model always identifies causal predictors with greater precision. Finally, we show an application of the proposed methodology to predict three externalizing behavorial scores in the combined QLSCD and QNTS longitudinal cohorts. 
\end{abstract}

\section{Introduction}

Our study of penalized generalized linear mixed models (GLMMs) for longitudinal traits was motivated by analyses of data collected from participants in the Quebec Longitudinal Study of Child Development (QLSCD) and the Quebec Newborn Twin Study (QNTS) to identify important genetic predictors for emotional and behavioral difficulties in childhood and adolescence, including externalizing (e.g., aggression) problems. Because of the longitudinal nature of the study, one needs to explicitly model the correlation between repeated measurements within an individual, one possibility being through the use of mixed-effects regression models. Moreover, genetic correlation between pairs of twins needs to be accounted for via a polygenic random effect~(\cite{Yang2011}), otherwise the study may be prone to a loss of power and spurious associations~(\cite{Yu2005,Price2010}). The generalized linear mixed model association test (GMMAT) proposed by~\citet{Chen2016} allows to include a known kinship matrix when analysing family samples with known pedigree structures in a homogeneous population, or an empirical genetic relatedness matrix (GRM) to account for both population structure and cryptic relatedness, for genome-wide association studies (GWAS) of continuous and binary traits. In addition, the authors have implemented random intercept only models, and random intercept and random slope models to account for random individual effects not attributable to the similarity between individuals. Typically, between-individuals similarity can be caused by genetic relatedness, shared environmental exposure or study sampling design.

Given the relative low sample sizes of the QLSCD ($n=721$) and QNTS cohort ($n=636$), a GWAS may fail to discover significant genetic variants that are associated with emotional and behavioral difficulties in childhood and adolescence. Moreover, obtaining effect size estimates for a large number of individual predictors via logistic or linear mixed models is highly computationally intensive using the GMMAT model, given that a new mixed model needs to be fitted for each predictor. This complicates the calculation of a polygenic score (PS) for longitudinal outcomes, in which variants effects across the genome are aggregated in order to predict complex traits~(\cite{Dudbridge2013,Choi2020}). Indeed, there is great clinical interest in being able to predict externalizing scores with precision as children following high-chronic trajectories of externalising and internalising behaviours have been shown to be at risk of negative long-term outcomes, including peer victimisation (\cite{vanLier2012,Oncioiu2020}), suicidal ideation and attempt (\cite{Orri2019,Forte2019}), and substance use~(\cite{Lemyre2018,Navarro2020,Zdebik2019}). 

Penalized models have been proposed as an alternative method to increase the power for identifying weaker genome-wide associations and interactions compared to univariable methods~(\cite{Chu2020,Li2010,Zhou2010,Wu2009,StPierre2023}). 
In this paper, we propose a lasso penalized mixed model framework for continuous and binary traits that allows the inclusion of more than one random effect to account for random individual effects not attributable to the genetic similarity between individuals. We study the performance of the average information restricted maximum likelihood (AIREML~(\cite{Gilmour1995})) algorithm when analyzing simulated data with both population structure and subjects relatedness for continuous and binary traits. In addition, we show that replacing the GRM by a sparse matrix greatly reduces the computational time required to fit the penalized model, while having little impact on the performance of the model in retrieving important predictors. Next, we compare the performance of our proposed model in retrieving important predictors for both continuous and binary traits, and demonstrate that it achieves better precision than the GMMAT model and that of a lasso penalized model without any random effect. Finally, we apply our proposed method for predicting externalizing scores in children from the combined QLSCD and QNTS cohorts and compare the performance of the lasso and adaptive lasso mixed models with respect to the predicted scores accuracy and models sparsity.

\section{Methods}
\subsection{Model}
Assume $y_{ij}$, $i=1,...,m$, $j=1,...,n_i$, is the measurement of a continuous or binary phenotype at time $t_{ij}$ for subject $i$, where $n=\sum_{i=1}^m n_i$ is the total number of observations. Let $\mathbf{C}_{ij}$ be a $1\times c$ row vector of possibly time-varying covariates for subject $i$, $\mathbf{G}_i$ a $1 \times p$ row vector of biallelic single nucleotide polymorphisms (SNPs) taking values $\{0,1,2\}$ as the number of copies of the minor allele, $(\bm{\theta}^\intercal,\bm{\beta}^\intercal)^\intercal$ a $(c + p) \times 1$ column vector of fixed covariate and additive genotype effects including the intercept.
We assume that $\mathbf{b}_0=(b_{01},...,b_{0m})^\intercal \sim \mathcal{N}(0, \sum_{k=1}^K \tau_k\mathbf{V}_k)$ is an $m \times 1$ column vector of random intercepts, $\bm{\tau}=(\tau_1, ...,\tau_K)^\intercal$ are the variance component parameters that account for the relatedness between individuals, and $\mathbf{V}_1,...,\mathbf{V}_K$ are known relatedness matrices. We typically define $\mathbf{V}_1$ as the GRM between individuals. Further, we assume that $\mathbf{b}_{1i} = (b_{11i}, b_{12i}, ..., b_{1ri})^\intercal\sim\mathcal{N}(0, \mathbf{D}(\mathbf \psi))$ is the $r\times 1$ column vector of subject-specific random effects for $i=1,...,m$ to account for the correlation between repeated measurements, where $\mathbf{D}$ is an $r \times r$ covariance matrix and $\bm{\psi}$ contains the unique elements of $\mathbf D$. Let $\mathbf{Z}_{ij}=(Z_{ij1}, ..., Z_{ijr})$ be a $1\times r$ covariate vector for subject-specific random effects $\mathbf{b}_{1i}$, possibly containing a non-polygenic random intercept and more than one random slope. The phenotypes $y_{ij}$'s are assumed to be conditionally independent and identically distributed given $(\mathbf{C}_{ij}, \mathbf{G}_i, \mathbf{Z}_{ij}, b_{0i}, \mathbf{b}_{1i})$ and follow any distribution with canonical link function $g(\cdot)$, mean $\E(y_{ij} | \mathbf{C}_{ij}, \mathbf{G}_i, \mathbf{Z}_{ij}, b_{0i}, \mathbf{b}_{1i}) =\mu_{ij}$ and variance $\Var(y_{ij}| \mathbf{C}_{ij}, \mathbf{G}_i, \mathbf{Z}_{ij}, b_{0i}, \mathbf{b}_{1i}) = \phi w_{ij}^{-1} \nu(\mu_{ij}),$ where $\phi$ is a dispersion parameter, $w_{ij}$ are known weights and $\nu(\cdot)$ is the variance function. We have the following GLMM for longitudinal data
\begin{align}\label{eq:model}
g(\mu_{ij}) &= \eta_{ij} = \mathbf{C}_{ij}\mathbf\theta + \mathbf{G_i} \bm{\beta} + b_{0i} + \mathbf{Z}_{ij} \mathbf{b}_{1i}.
\end{align}

We assume that the random effects vector $\mathbf{b}_0$ is independent of $\mathbf{b}_1=(\mathbf{b}_{11}^\intercal, ...,\mathbf{b}_{1r}^\intercal)$, such that the stacked random individual effects vector is $\mathbf{b}=(\mathbf{b}_0^\intercal, \mathbf{b}_{11}^\intercal,...,\mathbf{b}_{1r}^\intercal)^\intercal\sim N(0, \textrm{diag}\left\{\sum_{k=1}^K \tau_k\mathbf{V}_k , \mathbf{D} \otimes \mathbf{I}_m  \right\})$, where $\otimes$ is the Kroneker product. Typically, when there are no time trends and observations for the same individual are assumed to be exchangeable, a model with two random intercepts is appropriate, where the first random intercept captures the correlation induced by genetic relatedness, and the second intercept captures the correlation between repeated measurements. In the case where we observe or suspect individual-specific time trends to vary substantially, we can add one or more random slope to the model. 

For ease of presentation, let $\mathbf{y}=(y_{11},...,y_{1n_1}, ..., y_{m1},...,y_{mn_m})^\intercal$ be the stacked outcome vector and $\tilde{\mathbf{Z}}_k$ be an $n\times m$ block-diagonal matrix for the $k^{th}$ subject-specific random effect $\mathbf{b}_{1k}$ for $k=1,...,r$. For example, if $\mathbf{b}_{11}$ is a random intercept and $\mathbf{b}_{12}$ is a random slope, we would have for $m=3$ and $n_1=n_2=n_3=2$, $\tilde{\mathbf{Z}}_1 = \begin{bmatrix} 1 & 0 & 0\\ 1 & 0 & 0 \\ 0 & 1 & 0 \\ 0 & 1 & 0 \\ 0 & 0 & 1 \\ 0 & 0 & 1 \end{bmatrix}$ and $\tilde{\mathbf{Z}}_2 = \begin{bmatrix} Z_{112} & 0 & 0\\ Z_{122} & 0 & 0 \\ 0 & Z_{212} & 0 \\ 0 & Z_{222} & 0 \\ 0 & 0 & Z_{312} \\ 0 & 0 & Z_{322} \end{bmatrix}$. Next, let $\tilde{\mathbf{Z}}=\left[\tilde{\mathbf{Z}}_1, \tilde{\mathbf{Z}}_2, ...,\tilde{\mathbf{Z}}_r\right]$ be a $n\times mr$ block matrix, and $\mathbf{L}^\intercal=(\underbrace{\mathbf{L}_{1}^\intercal,...,\mathbf{L}_{1}^\intercal}_{n_1\ times},...,\underbrace{\mathbf{L}_{m}^\intercal,...,\mathbf{L}_{m}^\intercal}_{n_m\ times})$ be an $m\times n$ matrix of indicators such that $b_{0i}=\mathbf{L}_{i}\mathbf{b}_0$. Thus, for a model with two random intercepts and one random slope, the random effects for all observations is given by ${\mathbf{Lb}_0 + \tilde{\mathbf{Z}}\mathbf{b}_1 \sim N(0,\mathbf{L}\left(\sum_{k=1}^K\tau_k\mathbf{V}_k\right)\mathbf{L}^\intercal + \tilde{\mathbf{Z}}\{\begin{pmatrix} \psi_1 & \psi_2\\ \psi_2 & \psi_3 \end{pmatrix} \otimes \mathbf{I}_m\}\tilde{\mathbf{Z}}^\intercal)}$.

\citet{Chen2016} proposed a different variance-covariance structure for the random effects for all observations in model \eqref{eq:model}. They assumed that $\mathbf{b}_0 \sim \mathcal{N}(0, \sum_{k=1}^K \tau_k\mathbf{V}_k + \tau_{K+1}\mathbf{I}_m)$, $Cov(\mathbf{b}_0, \mathbf{b}_1)= \sum_{k=1}^K \tau_{K+1+k}\mathbf{V}_k + \tau_{2K+2}\mathbf{I}_m $ and $\mathbf{b}_1 \sim \mathcal{N}(0, \sum_{k=1}^K \tau_{2K+2+k}\mathbf{V}_k + \tau_{3K+3}\mathbf{I}_m)$. Thus, in the model with one ($K=1$) similarity matrix, they assumed that the random effects for all observations is given by $\mathbf{Lb}_0 + \tilde{\mathbf{Z}}\mathbf{b}_1 \sim N(0, \tilde{\mathbf{Z}}\{\begin{pmatrix} \tau_1 & \tau_3 \\ \tau_3 & \tau_5 \end{pmatrix} \otimes \mathbf{V}_1 + \begin{pmatrix} \tau_2 & \tau_4 \\ \tau_4 & \tau_6 \end{pmatrix} \otimes \mathbf{I}_m\}\tilde{\mathbf{Z}}^\intercal).$ Adding an additional random slope to model \eqref{eq:model} to account for extra sources of variability would increase the number of variance components to estimate from 6 to 12, compared to 7 variance components in our proposed variance-covariance structure, impacting not only the computational requirements to fit such model, but also the model interpretability. This is why we decided to adopt a different approach, where we assume that $\mathbf{b_0} \perp \mathbf{b}_1$ and that only the polygenic random intercept variance is proportional to the GRM. 

\subsection{Estimation}
In order to estimate the model parameters $\left(\bm{\theta}^\intercal,\bm{\beta}^\intercal,\phi, \bm{\tau}, \bm{\psi}\right)$ and perform variable selection, we use an approximation method to obtain an analytical closed form for the marginal likelihood of model \eqref{eq:model}. We propose to fit \eqref{eq:model} using a penalized quasi-likelihood (PQL) method, where the log integrated quasi-likelihood function is equal to
\begin{align}\label{eq:A3}
\ell_{PQL}(\bm{\Theta}, \phi, \bm{\tau}, \bm{\psi};\tilde{\mathbf{b}}) &= -\frac{1}{2}\text{log}\left|(\tilde{\mathbf{Z}}(\mathbf{D} \otimes \mathbf{I}_m )\tilde{\mathbf{Z}}^\intercal + {\mathbf{L}}(\sum_{k=1}^K\tau_k\mathbf{V}_k){\mathbf{L}}^\intercal)\mathbf{W} + \mathbf{I}_n\right| + \sum_{i,j} ql_{ij}(\bm{\theta};\tilde{\mathbf{b}}) \nonumber \\ 
&\qquad - \frac{1}{2}  \tilde{\mathbf{b}}^\intercal\left(\textrm{diag}\left\{\sum_{k=1}^K \tau_k\mathbf{V}_k , \mathbf D \otimes \mathbf{I}_m\right\}\right)^{-1}\tilde{\mathbf{b}},
\end{align} and $\bm{\Theta}=\left(\bm{\theta}^\intercal,\bm{\beta}^\intercal\right)^\intercal$, $\mathbf{W} = \phi^{-1}\mathbf{\Delta}^{-1}=\phi^{-1}\textrm{diag}\left\{ \frac{a_{ij}}{\nu(\mu_{ij})[g'(\mu_{ij})^2]}\right\}$ is a diagonal matrix containing weights for each observation, $a_{ij}$ are known weights, $ql_{ij}(\bm{\Theta};\mathbf{b}) = \int_{y_{ij}}^{\mu_{ij}}\frac{a_{ij}(y_{ij}-\mu)}{\phi\nu(\mu)} d\mu$ is the quasi-likelihood for the $jth$ observation from the $ith$ individual given the random effects $\mathbf b$, and $\tilde{\mathbf{b}}$ is the solution which maximizes $\sum_{i,j}^{} ql_{ij}(\bm{\Theta},\mathbf{{b}}) - \frac{1}{2}  \mathbf{{b}}^\intercal\left(\textrm{diag}\left\{\sum_{k=1}^K \tau_k\mathbf{V}_k , \mathbf D \otimes \mathbf{I}_m\right\}\right)^{-1}\mathbf{{b}}$.

In typical genome-wide studies, the number of genetic predictors is much greater than the number of observations ($p > n$), and the fixed effects parameter vector $\bm{\Theta}$ becomes unidentifiable when modelling $p$ SNPs jointly. 
Thus, we propose to add a lasso penalty~(\cite{lasso}) to the negative quasi-likelihood function in \eqref{eq:A3} to seek a sparse subset of genetic effects that gives an adequate fit to the data. We define the following objective function $Q_{\lambda}$ which we seek to minimize with respect to $(\bm{\Theta}, \phi, \bm{\tau}, \bm{\psi})$:
\begin{align}\label{eq:objfunc}
Q_{\lambda}(\bm{\Theta}, \phi, \bm{\tau}, \bm{\psi};\tilde{\mathbf{b}}) := -\ell_{PQL}(\bm{\Theta}, \phi, \bm{\tau}, \bm{\psi};\tilde{\mathbf{b}}) + \lambda\sum_j \nu_j|\beta_j|,
\end{align}
where $\lambda>0$ controls the strength of the overall regularization and $\nu_j$ are penalization weights that allows incorporating a priori information about the SNP effects. For example,~\citet{Zou2006} proposed the adaptive lasso where they defined the weights $\hat{\nu}_j=|\hat{\beta}_j|^{-\gamma}$ for $j=1,...,p$, and $\hat{\beta}_j$ is a root-$n$ consistent estimator of $\beta_j$, for example the ordinary least squares (OLS) estimator, and $\gamma>0$ is an additional tuning parameter.


\subsection{Estimation of variance components}
Jointly estimating the variance components and scale parameter vector $(\bm{\psi},\bm{\tau}, \phi)$ with fixed effects parameters vector $\bm{\Theta}$ is a computationally challenging non-convex optimization problem. Thus, as detailed in~\citet{StPierre2023}, we propose a two-step method where variance components and scale parameter are estimated only once under the null association of no genetic effect, that is assuming $\bm{\beta}=0$, using the AI-REML algorithm. Updates for $\bm{\tau}$, $\phi$ and $\bm{\psi}$ based on the AI-REML algorithm or a majorization-minimization algorithm~(\cite{Zhou2019}) requires iteratively inverting the $n\times n$ covariance matrix $\mathbf{\Sigma}$, with complexity $O(n^3)$. To reduce the computational cost of inverting the matrix $\mathbf \Sigma$, we can use the Woodbury matrix identity, and define the matrix $\mathbf{R} = \tilde{\mathbf{Z}}(\mathbf{D} \otimes \mathbf{I}_m )\tilde{\mathbf{Z}}^\intercal + \mathbf{W}^{-1}$, which yields
\begin{align*}
\mathbf{\Sigma}^{-1} = \mathbf{R}^{-1} - \mathbf{R}^{-1}\mathbf{L}(\mathbf{L}^\intercal\mathbf{R}^{-1}\mathbf{L}+(\sum\tau_k\mathbf{V}_k)^{-1})^{-1}\mathbf{L}^\intercal\mathbf{R}^{-1}.
\end{align*}
Because $\mathbf{R}$ and $\mathbf{L}^\intercal\mathbf{R}^{-1}\mathbf{L}$ are respectively block-diagonal and diagonal matrices, the complexity of inverting the $n \times n$ matrix $\mathbf{\Sigma}$ is similar to that of inverting the $m \times m$ matrix $\sum\tau_k\mathbf{V}_k$. To further reduce the computational complexity of the AI-REML estimation procedure when $\sum\tau_k\mathbf{V}_k = \tau_1\mathbf{V}_1$, that is when $K=1$, we propose replacing $\mathbf{V_1}$ by a sparse GRM~(\cite{jiang2019}), where pair-wise relatedness coefficients that are smaller than $2^{-9/2}$ are set to 0. This corresponds to a 3rd degree kinship threshold, meaning that anyone less related than first cousins are assumed to be unrelated. By rearranging the sparse GRM as a block-diagonal matrix, where each block consists of clusters of relatives, we show in the simulation study that it greatly reduces the computational time required to fit both the null and penalized models, and that the resulting bias in the variance components estimates has a limited impact on the performance of the penalized model to retrieve important predictors.

\section{Simulation study}
\subsection{Simulation model}

We performed simulation studies sampling real genotype data from a high quality harmonized set of 4,097 whole genomes from the Human Genome Diversity Project (HGDP) and the 1000 Genomes Project (1000G)~(\cite{Koenig2023}), including both related and unrelated individuals from seven distinct population groups (Table \ref{tab:pop}). At each of the 50 replications, we sampled {$10,000$} candidate SNPs from chromosome 21 and randomly selected 100 ($1\%$) to be causal. Let $S$ be the set of candidate causal SNPs, with $|S|=100$, then the causal SNPs fixed effects $\beta_s$ were generated from a Gaussian distribution $\mathcal{N}(0,h^2_S\sigma^2/|S|)$, where $h^2_S$ is the fraction of variance that is due to total additive genetic fixed effects and $\sigma^2$ is the total phenotypic variance. 

 \begin{table}[h]
\small\sf\centering
\caption{Number of samples by population for the high quality harmonized set of 4,097 whole genomes from the Human Genome Diversity Project (HGDP) and the 1000 Genomes Project (1000G).}\label{tab:pop}
\begin{tabular}{lccl}
  \hline
  Population & 1000 Genomes & HGDP & Total \\ 
  \hline
  African & 879 (28\%) & 110 (12\%) & 989 (24\%) \\ 
  Admixed American & 487 (15\%) &  62 (7\%) & 549 (13\%) \\ 
  Central/South Asian & 599 (19\%) & 184 (20\%) & 783 (19\%) \\ 
  East Asian & 583 (18\%) & 234 (25\%) & 817 (20\%) \\ 
  European & 618 (20\%) & 153 (16\%) & 771 (19\%) \\ 
  Middle Eastern &   0 & 158 (17\%) & 158 (4\%) \\ 
  Oceanian &   0 &  30 (3\%) &  30 (1\%) \\ \hline 
  Total & 3,166 & 931 & 4,097 \\ 
  Unrelated individuals & 2,520 & 880 & 3,400 \\ 
   \hline
\end{tabular}
\end{table}


We simulated a polygenic random intercept $\mathbf{b}_0\sim\mathcal{N}(0, h^2_{g}\sigma^2\mathbf{V}_1)$ where $h^2_{g}$ is the fraction of variance explained by the polygenic random effect and $\mathbf{V}_1$ is the estimated GRM using the PC-Relate method~(\cite{Conomos2016}). The polygenic random intercept $\mathbf{b}_0$ leverages the existing genetic relatedness in the sample due to familial or cryptic relatedness to simulate correlated phenotypes between individuals who share recent common ancestors. More specifically, pair-wise kinship coefficients were first estimated on a set of $13,750$ SNPs selected after LD pruning using the KING-Robust algorithm which is robust to population strucure~(\cite{Manichaikul2010}). Then, PCs were estimated using the PC-AiR method (principal components analysis in related samples) that allows to identify a diverse subset of mutually unrelated individuals such that the top PCs are constructed to only reflect the ancestry and to be robust to both known or cryptic relatedness in the sample~(\cite{Conomos2015}). Finally, the GRM was constructed from pair-wise kinship coefficients estimated using the residuals of a linear regression model after adjusting for the ancestry PCs calculated in the previous step. Hence, the PC-Relate method divides genetic correlations among sampled individuals into a component which represents familial relatedness, and another component which represents population structure~(\cite{Conomos2016}). To induce additional confounding due to population stratification, we simulated different intercepts $\pi_{0k}$, $k=1,..,7$, for each population in Table \ref{tab:pop} using a $U(0.1, 0.3)$ distribution.

For all individuals, we used the sex covariate available from the data set, and we simulated five measurements for age using a Normal distribution, after which 1 to 5 measurements were uniformly sampled to allow different number of observations per individual. To generate correlated observations for each individual, we simulated one random intercept, one random slope for the effect of age and one additional random slope representing the effect of a time-varying environmental exposure from a Gaussian distribution $\mathbf{b}_{1i} \sim \mathcal{N}(0, \mathbf{D})$ with the covariance matrix $\mathbf{D}$ equal to $\begin{bmatrix} 0.4 & -0.2 & 0.1 \\ -0.2 & 0.5 & 0.2 \\ 0.1 & 0.2 & 0.3\end{bmatrix}$. Then, for $i=1,...,4097$, $j=1,...,n_i$, continuous phenotypes were generated using the following model
\begin{align}\label{eq:simlogit1}
y_{ij} = \text{logit}(\pi_{0k}) -\text{log}(1.3)\times Sex_i+\text{log}(1.05)\times Age_{ij} + \sum_{s=1}^{100}\beta_s\cdot G_{is} + b_{0i} + \mathbf{Z}_{ij} \mathbf{b}_{1i} + \epsilon_{ij},
\end{align}
where $G_{is}$ is the (standardized) number of alleles for the $s^{th}$ causal SNP, $\mathbf{Z}_{ij}$ is the covariate vector for subjects-specific random effects $\mathbf{b}_{1i}$, and $\epsilon_{ij}$ is an error term from a standard normal distribution $\mathcal{N}(0, \phi)$. To simulate binary traits, we used a cutoff value $c$ determined such that $\P(y_{ij}^{01}=1) = \P(y_{ij} > c) = 0.2$.

\subsection{Results}
\subsubsection{Estimation of variance components}
\begin{figure}[]
\centering
\caption{Relative bias of variance parameters estimated under the null model of no genetic association when simulating continuous phenotypes with no causal predictor.}
\includegraphics[scale=0.6]{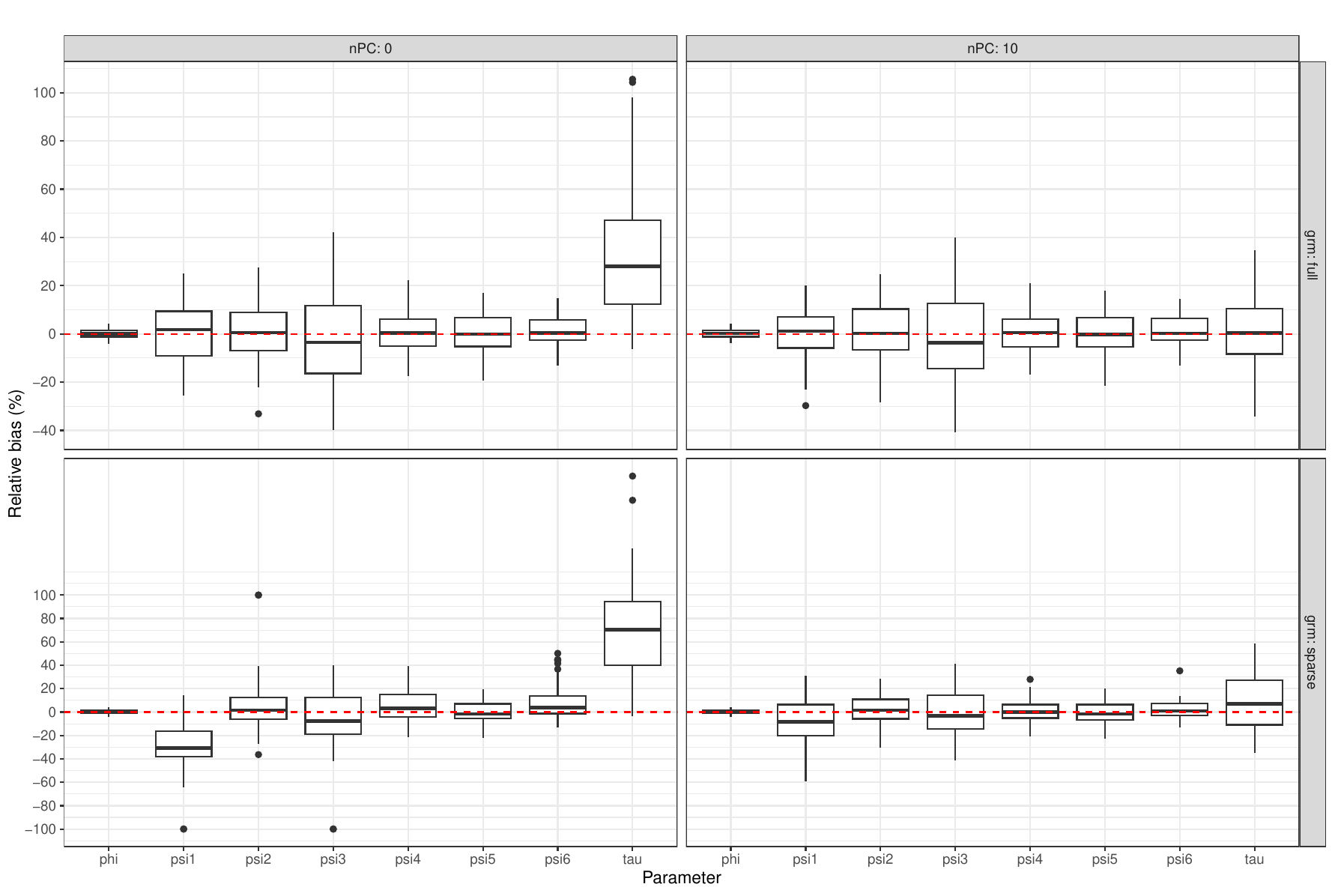}
\label{fig:1}
\end{figure}

We first simulated continuous phenotypes under the null model of no genetic association to study the impact on the variance components estimation procedure of including or not the first 10 PCs to control for genetic ancestry, as well as the impact of using a sparse GRM versus using the full GRM. In the upper-right panel of Figure \ref{fig:1}, we see that the median relative bias for all variance parameters is close to zero when fitting the model with the full GRM and the first 10 PCs, with the interquartile range (IQR) below $\pm 10\%$ for all parameters, except for the covariance parameter $\psi_3$. When replacing the full GRM by a sparse GRM (lower-right), the median relative bias for the variance of the non polygenic random intercept $\psi_1$ is around $-10\%$, while the median relative bias for the variance component of the polygenic random intercept $\tau$ is around $10\%$. Thus, the non-polygenic random intercept is capturing some of the variability that is not captured by the polygenic random intercept due to the fact that we are using a sparse GRM. The impact of using no PC to adjust for population structure in the sample is observed in the upper-left panel of Figure \ref{fig:1}, where the median relative bias for the variance component of the polygenic random intercept $\tau$ is around $30\%.$ This is explained by the fact that the PC-Relate estimator of the kinship coefficients is constructed after adjusting for PCs as predictors in linear regression models, and, thus, the residuals are orthogonal to the PCs. Genetic similarities due to more distant ancestry are therefore accounted for by using the PCs as covariates in the null model. When using only a sparse GRM without any PCs to adjust for both population structure and closer relatedness (lower-left), the median relative bias for $\tau$ and $\psi_1$ are quite important, as expected. Finally, in all four modelling strategies, the relative bias for the dispersion parameter $\phi$ which corresponds to the residual variance term of the errors $\epsilon_{ij}$ remains low. Adding the first 20 PCs to adjust for population structure decreased the mean relative bias for the variance component of the polygenic random intercept by a small amount compared to when using the first 10 PCs only (Supplementary Table \ref{supptab:1}). Results for binary traits (Supplementary Table \ref{supptab:3} and Supplementary Figure \ref{suppfig:1}) are consistent to what is observed for continuous traits, that is adjusting for population structure with the first 10 PCs reduces the relative bias of variance components, and the use of a sparse GRM slightly increases the relative biases compared to using a full GRM. Compared to simulations of continuous phenotypes, relative biases for variance components are more important in simulations with binary traits. 

\begin{figure}[]
\centering
\caption{Relative bias of variance parameters estimated under the null model of no genetic association when simulating continuous phenotypes with 100 causal predictors explaining 2\% of heritability.}
\includegraphics[scale=0.6]{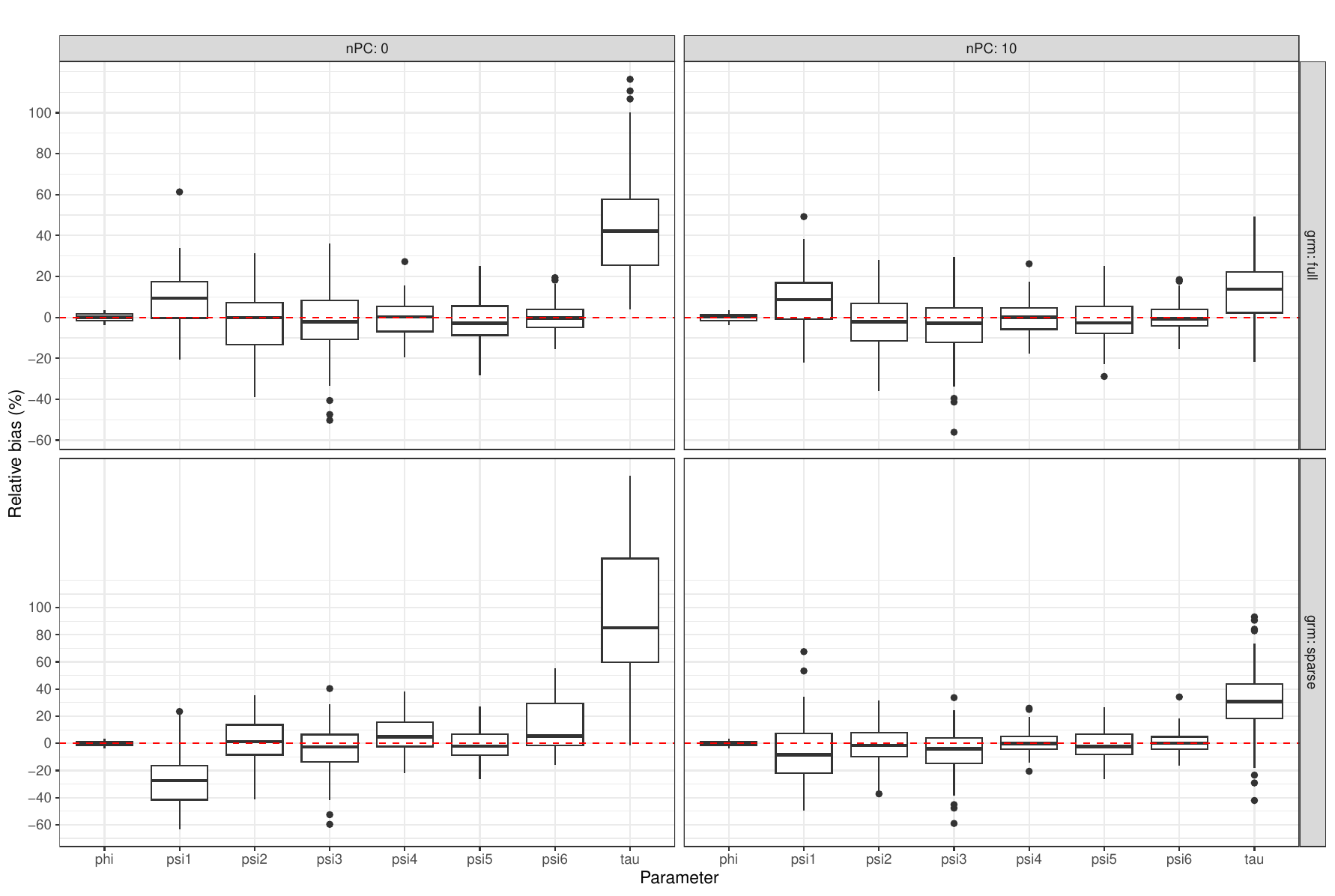}
\label{fig:2}
\end{figure}

We then simulated continuous phenotypes with 100 causal predictors explaining 2\% of heritability to assess the impact on the estimation of variance components when the model is misspecified, since under our proposed model, we estimate all variance parameters under the null model of no genetic association. As can be seen in Figure \ref{fig:2}, estimates of the variance parameters for the two random intercepts are upwardly biased even when using the full GRM with the first 10 PCs to account for population structure. When replacing the full GRM by a sparse one, the median relative bias for the variance of the non polygenic random intercept $\psi_1$ remains around $-10\%$, which is similar to when the null model is the true model (Figure \ref{fig:1}). However, we observe an increased relative bias when estimating the variance component of the polygenic random effect $\tau$, with a median value lying above $30\%$, compared to a median relative bias of $10\%$ when the model is not misspecified. Again, when fitting the model with either the full or sparse GRM but without any PC to control for population structure, estimates of variance parameters for the two random intercepts have large relative biases. Adding the first 20 PCs to adjust for population structure decreased the mean relative bias for the variance component of the polygenic random intercept by a small margin compared to when using the first 10 PCs only (Supplementary Table \ref{supptab:2}). Results for binary traits (Supplementary Table \ref{supptab:4} and Supplementary Figure \ref{suppfig:2}) are again consistent to results for continuous traits, albeit with relative biases that are larger in magnitude.

We present in Table \ref{tab:2} the computation time in minutes to fit the AI-REML algorithm when using a full GRM versus a sparse GRM for the simulation model with no causal predictor and for the simulation model with 100 causal predictors explaining 2\% of heritability. We see that although using a sparse GRM results in biased estimates of the variance parameters for the two random intercepts, the computation time is reduced by a factor of five for continuous phenotypes and by a factor of four for binary phenotypes.

\begin{table}[]
\caption{Computation time in minutes to fit the AI-REML algorithm as a function of the modelling strategy for the simulation model with no causal predictor (0\% heritability) and for the simulation model with 100 causal predictors explaining 2\% of heritability for both continuous and binary phenotypes. We present the median value with IQR in brackets.}
\label{tab:2}
\centering
\begin{tabular}{ll|cccc}
\hline
& & \multicolumn{2}{c}{Binary phenotypes} & \multicolumn{2}{c}{Continuous phenotypes} \\
Number of PCs & GRM & 0\% & 2\% & 0\% & 2\% \\ 
  \hline
  0 & full & 9.9 (7.0) & 8.3 (7.0) & 7.7 (0.4) & 6.9 (1.0) \\ 
     & sparse & 2.7 (1.9) & 2.5 (0.9) & 1.3 (0.2) & 1.4 (0.1) \\ 
   10 & full & 7.3 (7.6) & 8.3 (2.6) & 7.4 (0.3) & 6.7 (1.1) \\ 
    & sparse & 1.6 (1.0) & 1.2 (1.2) & 1.1 (0.2) & 1.2 (0.1) \\ 
   20 & full & 9.0 (8.1) & 8.5 (4.4) & 7.5 (1) & 8.2 (1.2) \\ 
    & sparse & 1.6 (0.9) & 1.7 (1.2) & 1.3 (0.1) & 1.7 (0.1) \\ 
   \hline
\end{tabular}
\end{table}

\subsubsection{Selection of genetic predictors}

\begin{figure}[h]
\centering
\caption{Precision-recall curve for selection of genetic predictors for our proposed method as a function of the modelling strategy. The left and right panels illustrate the average performance of the method over 50 replications for the simulation model with continuous phenotypes and 100 causal predictors explaining 2\% and 10\% of heritability respectively.}
\includegraphics[scale=0.6]{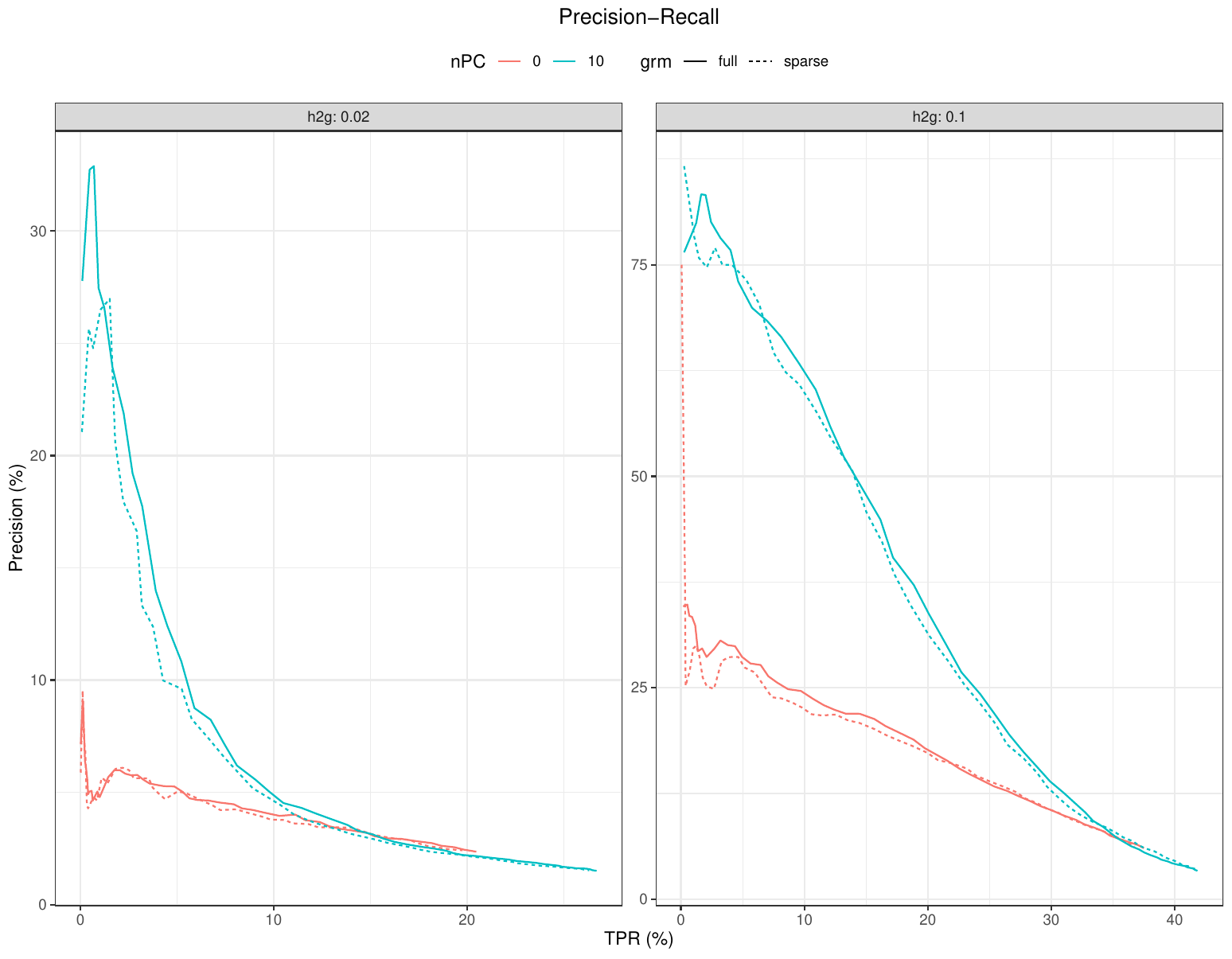}
\label{fig:3}
\end{figure}

We present in Figure \ref{fig:3} the Precision-Recall (PR) curve to illustrate the performance of our proposed method in retrieving causal genetic predictors as a function of the modelling strategy for the simulation model with continuous phenotypes and 100 causal predictors explaining $2\%$ and $10\%$ of heritability respectively. The PR curve displays the precision of a method, i.e. the proportion of selected predictors that are truly causal, as a function of the true positive rate (TPR), that is the percentage of true causal predictors that are selected by the model. Because the number of non-causal predictors is significantly greater than the number of causal predictors in genetic association studies, the PR curve is a more robust tool than the receiving operator characteristic (ROC) curve which is sensitive to the imbalance between the number of causal and non-causal predictors, since the false positive rate (FPR) is naturally weighted down due to the very large number of true negatives~(\cite{Saito2015}). We see that including the 10 PCs as covariates to adjust for population structure in the penalized model greatly increases the ability to retrieve causal predictors. On the other hand, using a sparse GRM in place of the full GRM to adjust for relatedness in both the null and penalized models has little impact on the performance of the model, which is encouraging. Results for binary phenotypes, presented in Supplementary Figure \ref{suppfig:3}, are consistent with results obtained when simulating continuous phenotypes. Finally, median computation times to fit the lasso regularization path for our proposed penalized mixed model is presented in Table \ref{tab:2.1}. Using a sparse GRM in place of the full GRM reduces the median computation time by at least a factor of two in all simulations. 

\begin{table}[]
\caption{Computation time in minutes to fit the lasso regularization path for each modelling strategy for the simulation model with no causal predictor (0\% heritability) and for the simulation model with 100 causal predictors explaining 2\% of heritability for both continuous and binary phenotypes. We present the median value with IQR in brackets.}
\label{tab:2.1}
\centering
\begin{tabular}{ll|cccc}
  \hline
& & \multicolumn{2}{c}{Binary phenotypes} & \multicolumn{2}{c}{Continuous phenotypes} \\
Number of PCs & GRM & 0\% & 2\% & 0\% & 2\% \\ 
  \hline
0 & full & 82.3 (17.6) & 88.0 (19.8) & 60.8 (18.6) & 66.8 (26.4) \\ 
   & sparse & 34.2 (10.5) & 35.1 (6.9) & 25.0 (15.2) & 24.2 (9.5) \\ 
  10 & full & 96.0 (14.8) & 97.2 (15.3) & 113 (29.4) & 81.5 (19.7) \\ 
   & sparse & 45.2 (8.6) & 40.0 (8.5) & 57.5 (16.3) & 35.1 (8.1) \\ 
   \hline
\end{tabular}
\end{table}

\begin{figure}[h]
\centering
\caption{Precision-recall curve for selection of genetic predictors for the three compared methods. The left and right panels illustrate the average performance with 95\% confidence interval of the methods over 50 replications for the simulation model with continuous phenotypes and 100 causal predictors explaining 2\% and 10\% of heritability respectively.}
\includegraphics[scale=0.55]{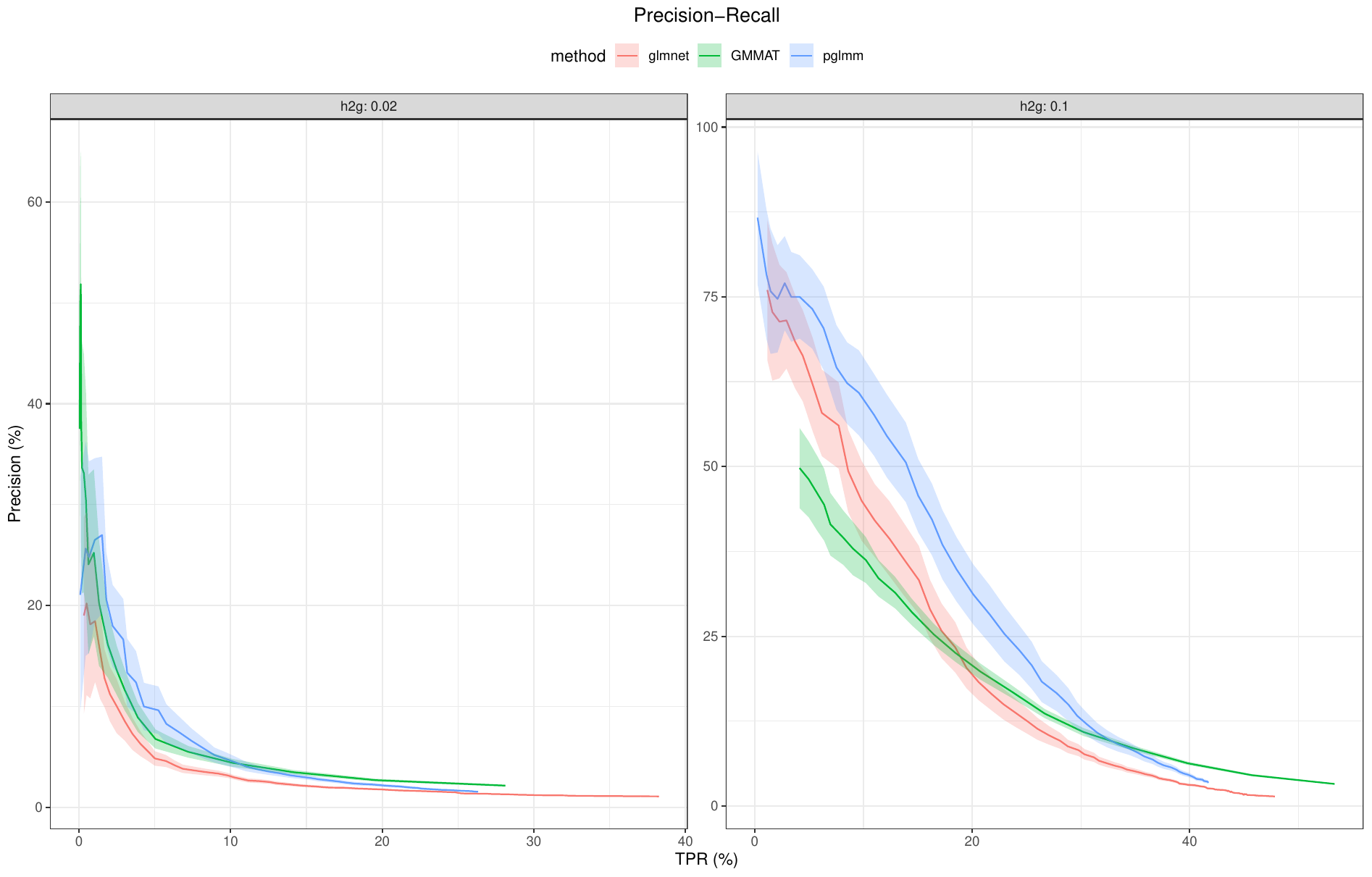}
\label{fig:4}
\end{figure}

Next we compared the performance of our method versus that of a standard lasso, using the \texttt{Julia} package \texttt{GLMNet} which wraps the \texttt{Fortran} code from the original \texttt{R} package \texttt{glmnet}~(\cite{glmnet}). The default implementation for \texttt{glmnet} and \texttt{pglmm} is to find the smallest value of the tuning parameter $\lambda$ such that no predictors are selected in the model, and then to solve the penalized minimization problem over a grid of decreasing values of $\lambda$. For these two methods, we used a grid of 100 values of $\lambda$ on the log10 scale with $\lambda_{min}=0.01\lambda_{max}$, where $\lambda_{max}$ is chosen such that no predictors are selected in the model. We also compared the performance of our method with that of GMMAT, using the freely available \texttt{R} package. Variable selection for GMMAT was performed by varying the cutoff value $p_\alpha$ such that all predictors with a p-value smaller than $p_\alpha$ were retained in the model. For GMMAT and our proposed method, we use a sparse GRM to account for relatedness between individuals. For all three methods, population structure was accounted for by adding the first 10 PCs as additional covariates. As can be seen on Figure \ref{fig:4}, our proposed method's ability to retrieve causal predictors is uniformly superior to that of both \texttt{glmnet} and GMMAT. Results for binary phenotypes, presented in Supplementary Figure \ref{suppfig:4}, are again consistent with those obtained when simulating continuous phenotypes.

\clearpage
\section{Identification of genetic predictors for emotional and behavioral difficulties in childhood and adolescence}

Emotional and behavioral difficulties in childhood and adolescence are associated with significant impairment in various domains of functioning and are a major public health concern~(\cite{Ogundele2018}). The etiology of such problems is complex and involves both genetic and environmental determinants that are likely to interact and change over the course of development. Understanding the genetic and environmental risk factors underlying these difficulties is crucial for the development of effective prevention and intervention strategies. However, most of the research in genetic epidemiology is conducted in samples of adults, and then generalized to children and adolescences. This is mainly due to the fact that samples to conduct GWAS in children and adolescent are of relatively small size, which is not adapted to the standard GWAS approaches. 

The objective of this study is to identify genetic predictors for emotional and behavioral difficulties in childhood and adolescents from the the Quebec Longitudinal Study of Child Development (QLSCD) and the Quebec Newborn Twin Study (QNTS). The QLSCD was designed to examine the long-term associations of preschool physical, cognitive, social, and emotional development with biopsychosocial development across childhood, adolescence, and young adulthood~(\cite{Orri2021}). Data have been collected annually or every 2 years from children born in 1997/1998 in the Canadian province of Quebec and followed up from ages 5 months to 25 years. Data were collected, in either English or French, by trained interviewers from the person most knowledgeable about the child (mother in $>98\%$ of the cases). The QNTS is a population-based cohort initially including 1324 twins (i.e., 662 pairs) born between April 1st, 1995 and December 31st, 1998 in the greater Montreal area of the province of Québec, Canada~(\cite{Boivin2012}). Zygosity was initially assessed via questionnaire and confirmed with DNA tests on a subsample of $n=123$ same-sex pairs ($96\%$ correspondence;~(\cite{ForgetDubois2003})). 

\subsection{Outcomes and covariates}
When children were 6, 7, 8, 9, 10, and 12 years of age, school teachers rated their social and emotional behavior in the past 6 months using validated questionnaires. For this study, we focused on three externalizing behaviors: aggression, hyperactivity and opposition. Hyperactivity, aggression and opposition externalizing behaviors scores were rated on a 3-point Likert scale (0=never/1=sometimes/2=often) using respectively six, ten and four validated items from the Social Behavior Questionnaire~\cite{Collet2022}, and then averaged at each age (range 0-2, higher scores indicating higher propensity to display such behavior). Children’s behavior was assessed by a different teacher each year, thus reducing risks of rater bias. Socio-economic status was derived using household income, education level and prestige of the profession of both parents, and then standardized with zero mean and unit standard deviation (SD)~(\cite{gouv}). We dichotomized the variable (Low vs High) using 0 as the cutoff value.

\subsection{Genotype data}
Genome-wide genotype data was available from $n=721$ participants in the QLSCD cohort and $n=641$ participants in the QNTS cohort. Biological samples from the QNTS participants were genotyped in two batches, the first in 2016 and the second in 2019. QLSCD participants were genotyped in 2016. For all batches, Illumina PsychArrayv1.X was used with assembly b37. Data were exported in FWD/REV format using Illumia’s GenomeStudio software. Quality control (QC) was conducted with PLINK v1.90b6.20 and R v4.0, and all steps are detailled in Appendix \ref{appendix:3}. For the calculation of ancestry components used to determine genetic outliers and as covariates in the analyses, pre-imputation genotype data were used. Genetic outliers were defined as individuals with a distance from the mean of $>4$ SD in the first eight multidimensional scaling (MDS) ancestry components. Additional variant filtering steps for calculation of ancestry components were the removal of variants with a MAF $<0.05$ or Hardy-Weinberg equilibrium (HWE) exact test p-value $<10^{-3}$ in sets of unrelated individuals, removal of variants mapping to the extended MHC region (chromosome 6, 25-35 Mbp) or to a typical inversion site on chromosome 8 (7-13 Mbp), and linkage disequilibrium (LD) pruning. Next, the pairwise identity-by-state (IBS) matrix of all individuals was calculated using filtered genotype data. MDS analysis was performed on the IBS matrix using the eigendecomposition-based algorithm in PLINK v1.90b6.7 (QLSCD) and PLINK v1.90b5.2 (QNTS). Imputation was conducted using SHAPEIT v2 (r837)~(\cite{Delaneau2013}), IMPUTE2 v2.3.2~(\cite{Howie2009}), and the 1000 Genomes Phase 3 reference panel. After imputation, variants with a MAF $<1\%$, an HWE exact test p-value $<1\times10^{-6}$, and an INFO metric $<0.8$ were removed. Finally, variants that were not SNPs or that were strand-ambiguous were removed, leaving about four millions imputed SNPs for the analysis. Imputed SNPs were converted into PLINK binary format.

\subsection{Methods}
\subsubsection{Imputation of missing data}
Socio-economic status was missing for 3 (0.42\%) participants from the QLSCD cohort and 110 (17\%) participants from the QNTS cohort. Thus, we imputed the missing covariate using the average in each cohort independently. A significant number of children whose genotype data were available had missing data for all time points (1\% of participants in the QLSCD cohort; 12\% of the participants in the QNTS cohort). In order to avoid discarding them from GWAS which may reduce the power to find any significant SNPs, we imputed their externalizing scores using the average score stratified by cohort, sex, binary socio-economic status and age. For children who were lost to follow-up, externalizing scores were imputed using the last observation carried forward (LOCF) procedure. In the following analyses, we compare the results between the complete case analysis (CCA, N = 1217) and the single imputation (SI, N = 1357) approach.

\subsubsection{GWAS}
We ran a GWAS using the GMMAT package in \texttt{R} using age, age$^2$, cohort, binary socio-economic status, sex and the first 10 PCs as fixed-effects covariates in the model. To account for the correlations between repeated measurements, we added a random intercept and a random slope for age. To account for the genetic correlation between individuals, we used a sparse GRM that was calculated using the PC-Relate method~(\cite{Conomos2016}). The GWAS analysis was performed on just over 4 millions imputed SNPs that passed QC.

\subsubsection{Prediction model}
We used our proposed lasso penalized mixed model for selecting important predictors and for building a prediction model for the externalizing behaviors. We used the same fixed-effects covariates in the model as for the GWAS. To account for the correlations between repeated measurements, we again added a random intercept and a random slope for age, and we used a sparse GRM to account for the genetic correlation between individuals. We fitted the model removing the last observation for each participant (train set), and used the last observation for each participant (test set) to assess the accuracy of the predicted externalizing scores. We compared the performance of the lasso penalized mixed model with an adaptive lasso mixed model, using the \texttt{bigstatsr} package in \texttt{R} to fit an elastic-net penalty model on the training data. More specifically, we used the estimated predictors coefficients from the elastic-net model to define the adaptive lasso weights to be used in our proposed penalized mixed model, such that for the $j^{th}$ predictor, the weight was defined as $\hat w_j = |{\hat{\beta}_j}^{enet}|^{-0.25}$.

To assess the calibration and predictive performance of the prediction models, we report a mean squared prediction error (MSPE)-based definition of the $R^2$ coefficient of determination~(\cite{Staerk2024}) defined as $$R^2_{MSPE} = 1 - \frac{\sum_{i=1}^m(y_i - \hat{y_i})^2}{\sum_{i=1}^m(y_i - \bar y)^2},$$ where $\hat{y_i}$ is the predicted score for the $i^{th}$ participant, for $i=1,...,m$, and $\bar y$ is the mean score in the test set. We note that if $R^2_{MSPE}$ is negative, this means that a simple intercept model on the test data performs better than the polygenic model in predicting externalizing scores. Compared to the squared correlation between predicted and observed values which only focuses on the discriminative performance of the model, $R^2_{MSPE}$ takes both discrimination and calibration of the prediction model into account. To reduce the number of predictors to incorporate in the penalized mixed models, we merged the list of imputed SNPs that passed QC ($\sim$ 4 millions SNPs) with the list of SNPs from the third phase of the International HapMap Project~(\cite{HapMap3, HapMap3_snps}), containing a total of 1.2 millions of SNPs that are generally well imputed in most studies. After merging both list of SNPs, the final analysis included a total of 735K SNPs to be included in our penalized regressions mixed models.

\subsection{Results}
Characteristics of the QLSCD and QNTS cohorts participants are presented in Table \ref{tab:3}. For both cohorts, externalizing scores are higher on average in males compared to females. Aggression mean scores are similar between the two cohorts, as opposed to hyperactivity and opposition mean score values that are higher in the QNTS cohort. We present in Table \ref{tab:4} the list of SNPs with a p-value smaller than $5\times10^{-8}$ for either the CCA or SI model, which is generally considered as the genome-wide significance threshold~(\cite{Peer2008}). P-values were obtained by fitting the GMMAT model on all imputed SNPs, using a score test statistics that was computed using variance components estimated under the null model of no genetic association. When two significant SNPs were in LD, defined as a squared correlation based on genotypic allele counts greater than 0.2, only the most significant was retained. A total of 9 SNPs were significant in the CCA model for the hyperactivity externalizing score, but none were significant in the SI model. Similarly, a total of 9 SNPs were significant in the CCA model for the aggression externalizing score, but none were significant in the SI model. Finally, 2 SNPs were significant in the GWAS of the opposition externalizing score for both analyses. We note that there was no overlap between the SNPs that were found to be significant among the three externalizing scores, which may be surprising given that the correlation coefficient between the three scores ranges from 0.57 to 0.70. 

The predictive performance of the lasso and adaptive lasso penalized mixed models as a function of the number of selected genetic predictors is presented in Supplementary Figures \ref{suppfig:5} and \ref{suppfig:6} respectively. For the lasso penalized mixed model, the best predictions for both imputation methods are obtained for the opposition externalizing score (CCA $R^2_{MSPE} = 0.310$ and SI $R^2_{MSPE} = 0.441$), followed by hyperactivity (CCA $R^2_{MSPE} = 0.304$ and SI $R^2_{MSPE} = 0.420$) and aggression (CCA $R^2_{MSPE} = 0.261$ and SI $R^2_{MSPE} = 0.403$) as presented in Table \ref{tab:5}. The number of selected genetic predictors in all models is relatively high, ranging from 601 to 851 for the CCA, and from 965 to 1163 for the models using SI to impute missing values.

For the adaptive lasso penalized mixed model, the best predictions for both imputation methods are again obtained for the opposition externalizing score (CCA $R^2_{MSPE} = 0.294$ and SI $R^2_{MSPE} = 0.439$), followed by hyperactivity (CCA $R^2_{MSPE} = 0.301$ and SI $R^2_{MSPE} = 0.411$) and aggression (CCA $R^2_{MSPE} = 0.235$ and SI $R^2_{MSPE} = 0.376$). Thus, the proportion of variance explained by the estimated polygenic score is slightly lower in the adaptive lasso compared to the lasso regularized prediction model. On the the other hand, the number of selected genetic predictors in all models is comparatively lower in the adaptive lasso mixed model, ranging from 419 to 628 for the CCA, and from 744 to 966 for the models using SI to replace missing values.

The set of overlapping SNPs, i.e. SNPs that were selected by the penalized mixed model for all three externalizing scores included 3 SNPs in the complete case analysis and 16 SNPs in the single imputation analysis as presented in Supplementary Table \ref{supptab:5}. A total of 2 SNPs in the complete case analysis and 17 SNPs in the single imputation analysis were selected by the adaptive penalized mixed model for all three externalizing scores as presented in Supplementary Table \ref{supptab:6}. We report for each SNP the gene where they are located when applicable, and if they are intronic, missense or non coding variants. 

Finally, we examined the impact of imputing missing data on the variance component estimates under our proposed model. As shown in Supplementary Table \ref{tab:varcomps}, the estimates for the polygenic variance component ($\hat\tau$) were consistent in both the CCA and the SI analysis. However, imputing missing data led to lower estimates for the residual variance parameter ($\hat\phi$) compared to CCA, while the estimates for the variance of the non-polygenic random intercept ($\phi_1$) were higher in the SI analysis. This result aligns with the imputation method we used, which involved imputing missing externalizing scores using stratified sample averages, thereby reducing variability between individuals. The increase in within-individual variability after imputing missing data suggests that differences between individuals play a significant role relative to within-individual fluctuations in explaining trajectories of externalizing behaviors.

\begin{table}[]
\small
\caption{Characteristics of the Quebec Longitudinal Study of Child Development (QLSCD) and the Quebec Newborn Twin Study (QNTS) participants.}
\label{tab:3}
\centering
\begin{tabular}{lcccccc}
\hline
& \multicolumn{6}{c}{Cohort} \\ \cline{2-4} \cline{5-7}
                           & \multicolumn{3}{c}{QLSCD} & \multicolumn{3}{c}{QNTS} \\ 
                           & Females & Males & Total & Females & Males & Total \\ \hline
n                        & 398 (55\%) & 323 (45\%) & 721                            & 319 (50\%) & 317 (50\%) & 636                             \\
Socio-economic status (\%) & & &    & & &   \\
\ \ \ \ High   & 206 (52\%) & 165 (51\%) & 371 (51\%)  & 133 (42\%) & 139 (44\%) & 272 (43\%)                      \\
Aggression score, Mean (SD) \\
\ \ \ \ \ \ \ \ 6 years old  & 0.16 (0.28) & 0.33 (0.45) & 0.23 (0.37)  & 0.18 (0.36) &  0.38 (0.54) & 0.28 (0.47) \\
\ \ \ \ \ \ \ \ 7 years old   & 0.15 (0.30) & 0.34 (0.40) & 0.23 (0.36) & 0.11 (0.29) &  0.43 (0.54) & 0.27 (0.46)  \\
\ \ \ \ \ \ \ \ 8-9 years old  & 0.14 (0.29) &  0.31 (0.40) & 0.22 (0.35) & 0.10 (0.30) & 0.37 (0.54) & 0.23 (0.45) \\
\ \ \ \ \ \ \ \ 10 years old   & 0.11 (0.23) &  0.32 (0.45) & 0.20 (0.36) &  0.11 (0.31) &  0.35 (0.50) & 0.23 (0.43) \\
\ \ \ \ \ \ \ \ 12 years old  & 0.09 (0.20) & 0.26 (0.39) & 0.16 (0.31) &  0.02 (0.10) &  0.25 (0.42) & 0.13 (0.32)  \\ \\
Hyperactivity score, Mean (SD) \\
\ \ \ \ \ \ \ \ 6 years old              & 0.27 (0.39) &  0.56 (0.56) & 0.39 (0.49)  & 0.41 (0.50) & 0.70 (0.65) &   0.55 (0.60) \\
\ \ \ \ \ \ \ \ 7 years old              & 0.28 (0.40) & 0.57 (0.56) & 0.41 (0.50)  & 0.36 (0.48) &  0.65 (0.61) &   0.51 (0.57) \\
\ \ \ \ \ \ \ \ 8 years old              & 0.28 (0.40) & 0.53 (0.52) & 0.39 (0.47)  &  0.57 (0.44) &  0.67 (0.44) &   0.62 (0.44)  \\
\ \ \ \ \ \ \ \ 9 years old              & - & - &     -    & 0.30 (0.46) & 0.60 (0.62)  &   0.45 (0.56)  \\
\ \ \ \ \ \ \ \ 10 years old             & 0.17 (0.34) & 0.50 (0.54) & 0.32 (0.47)  & 0.25 (0.41) &  0.56 (0.55) &   0.41 (0.51)  \\
\ \ \ \ \ \ \ \ 12 years old             & 0.12 (0.24) &  0.42 (0.48) & 0.25 (0.40)  & 0.21 (0.38) &  0.46 (0.52) &   0.33 (0.47)  \\ \\
Opposition score, Mean (SD) \\
\ \ \ \ \ \ \ \ 6 years old             & 0.19 (0.33) & 0.38 (0.49) &  0.27 (0.42) &  0.27 (0.45) &  0.45 (0.56) & 0.36 (0.51) \\
\ \ \ \ \ \ \ \ 7 years old              &  0.17 (0.34) &  0.40 (0.51)  & 0.27 (0.44) & 0.21 (0.38) &  0.46 (0.57) & 0.34 (0.50) \\
\ \ \ \ \ \ \ \ 8-9 years old            &  0.20 (0.40) & 0.41 (0.49) & 0.29 (0.45) & 0.23 (0.44) &  0.44 (0.56) & 0.33 (0.51) \\
\ \ \ \ \ \ \ \ 10 years old            &  0.18 (0.36) & 0.41 (0.55) & 0.28 (0.47) & 0.22 (0.41)  &   0.43 (0.53) & 0.33 (0.49) \\
\ \ \ \ \ \ \ \ 12 years old           &  0.15 (0.33) & 0.40 (0.51) & 0.26 (0.44) &  0.16 (0.28) & 0.41 (0.52) & 0.28 (0.43) \\
\hline
\end{tabular}
\end{table}

\begin{sidewaystable}[ht]
\smaller
\centering
\begin{threeparttable}
\caption{Top SNPs identified from GWAS of three externalizing scores in the combined QLSCD and QNTS cohortsf.}\label{tab:4}
\begin{tabular}{llllcclccccccc}
  \hline
Ext. & SNP & Chr & Position & A1 / A2 & MAF & Gene : Consequence & \multicolumn{2}{c}{N} & \multicolumn{2}{c}{p-value} & \multicolumn{2}{c}{$\hat{\beta}$} \\ 
  score \\ \hline
Hyp & & & & & & & CCA & SI & CCA & SI & CCA & SI \\ 
& rs144702376 & 5 & 119325132 & T / C & 0.014 & None & 1191 & 1329 & $4.28\times10^{-8}$ & $1.06\times10^{-5}$ & $2.8\times10^{-4}$ & $3.9\times10^{-4}$ \\ 
& rs115921653 & 6 & 10089651 & A / G & 0.017 & OFCC1 : Intronic & 1210 & 1348 & $2.38\times10^{-8}$ & $9.62\times10^{-6}$ & $2.8\times10^{-4}$ & $3.9\times10^{-4}$ \\
& rs149436805 & 6 & 97566274 & T / C & 0.015 & KLHL32 : Intronic &  1211 & 1350 & $2.90\times10^{-8}$ & $1.20\times10^{-4}$ & $1.9\times10^{-4}$ & $3.9\times10^{-4}$ \\
& rs77406243 & 8 & 89633141 & A / G & 0.010 & LOC105375630 : Intronic &  1211 & 1350 & $4.09 \times10^{-9}$ & - & $2.4\times10^{-4}$ & $3.9\times10^{-4}$ \\
& rs80133571 & 12 &    108969583 & C / T & 0.010 & None & 1204 &  1338 & 3.78$\times10^{-8}$ & 1.52$\times10^{-5}$ & $2.6\times10^{-4}$ & $3.9\times10^{-4}$ \\ 
& rs112641948 & 15 &     76967976 & T / C & 0.016 & SCAPER : Intronic & 1208 & 1347 & 4.98$\times10^{-8}$ & 5.43$\times10^{-7}$ & $2.6\times10^{-4}$ & $3.9\times10^{-4}$ \\
& rs8053367 &  16 & 53875484 & T / G & 0.468 & FTO : Intronic & 1216 & 1356 & $4.62\times10^{-8}$ & $4.53\times10^{-4}$ & $2.5\times10^{-4}$ & $3.9\times10^{-4}$ \\ 
& rs17231282 & 18 & 67692722 & T / C & 0.018 & RTTN : Intronic & 1194 & 1331 & 3.01$\times10^{-8}$ & 1.76$\times10^{-3}$ & $2.5\times10^{-4}$ & $3.9\times10^{-4}$ \\ 
& rs12971894 &  19 &  29985790 & A / G & 0.123 & VSTM2B-DT : Intronic & 1197 & 1336 & 3.58$\times10^{-8}$ & 3.28$\times10^{-4}$ & $2.4\times10^{-4}$ & $3.9\times10^{-4}$ \\
Aggr \\
& rs13393674 & 2 &    158403762 & T / C & 0.052 & ACVR1C : Intronic & 1193 &         1330 & 8.69$\times10^{-8}$ & 4.38$\times10^{-9}$ & 2.3$\times10^{-4}$ & 8.1$\times10^{-5}$ \\ 
&  rs10812324 & 9 &     25933433 & C / T & 0.047 & None & 1200 &         1338 & 4.06$\times10^{-8}$ & 1.08$\times10^{-4}$ & 2.6$\times10^{-4}$ & 8.0$\times10^{-5}$ \\ 
&  rs7911165 & 10 &    131626650 & C / T &   0.359 &   LOC107984281 : NC   & 1216 &         1357 & 4.09$\times10^{-8}$ & 2.34$\times10^{-5}$ & 2.2$\times10^{-4}$ & 7.8$\times10^{-5}$ \\ 
&  rs113352461 & 12 &     33390548 & G / A &  0.010 &   None    & 1199 &         1337 & 8.56$\times10^{-9}$ & 7.92$\times10^{-5}$ & 2.3$\times10^{-4}$ & 7.9$\times10^{-5}$ \\ 
&  rs202096 & 13 &     30820199 & A / G &   0.350 &   KATNAL1 : Intronic   & 1206 &         1346 & 1.19$\times10^{-8}$ & 1.42$\times10^{-5}$ & 2.1$\times10^{-4}$ & 8.0$\times10^{-5}$ \\ 
&  rs144975159 & 15 &     52666187 & T / G &  0.012 & MYO5A : Intronic & 1214 &         1355 & 4.26$\times10^{-8}$ & 2.59$\times10^{-6}$ & 2.3$\times10^{-4}$ & 7.9$\times10^{-5}$ \\ 
&  rs4792882 & 17 &     43869277 & G / T &   0.018 &   CRHR1 : Intronic   & 1210 &         1347 & 1.09$\times10^{-8}$ & 2.08$\times10^{-6}$ & 2.3$\times10^{-4}$ & 8.1$\times10^{-5}$ \\ 
&  rs187443053 & 18 &     67161236 & A / G &   0.012 &  DOK6 : Intronic    & 1212 &         1353 & 3.29$\times10^{-8}$ & 3.35$\times10^{-6}$ & 2.5$\times10^{-4}$ & 8.0$\times10^{-5}$ \\ 
&  rs77702389 & 21 &     32712872 & T / G &   0.049 &  TIAM1 : Intronic    & 1216 &         1357 & 2.97$\times10^{-12}$ & 6.48$\times10^{-7}$ & 2.2$\times10^{-4}$ & 8.0$\times10^{-5}$ \\ 
Opp \\
& rs116541675 & 2 &     18419626 & G / A &  0.011 &  None     & 1213 &         1354 & 3.42$\times10^{-8}$ & 2.00$\times10^{-9}$ & 2.9$\times10^{-4}$ & 6.7$\times10^{-4}$ \\ 
& rs145226968 & 3 &    153021039 & C / T &   0.011 &  None    & 1195 &         1333 & 2.87$\times10^{-7}$ & 1.74$\times10^{-8}$ & 3.0$\times10^{-4}$ & 6.8$\times10^{-4}$ \\
   \hline
\end{tabular}
\begin{tablenotes}
      \small
      \item MAF = Minor allele frequency. N = Number of non-missing observations. CCA = Complete case analysis. SI = Single imputation.
      \item Hyp = Hyperactivity. Aggr = Aggression. Opp = Opposition.
      \item NC = Non-coding.
      \item Note 1: MAF was reported for the effect allele A1.
      \item Note 2: p-values were obtained using the GMMAT score test. Effect size estimates were obtained using the GMMAT Wald test.
    \end{tablenotes}
\end{threeparttable}
\end{sidewaystable}

\begin{table}[]
\smaller
\centering
\begin{threeparttable}
\caption{Performance of the best prediction models on the test set for each externalizing score, imputation model and regularization procedure.}\label{tab:5}
\begin{tabular}{llcccccc}
\hline
  & & \multicolumn{3}{c}{Lasso mixed model} & \multicolumn{3}{c}{Adaptive lasso mixed model} \\
  Externalizing & & Number of selected & & & Number of selected   \\ 
  score & Analysis & genetic predictors & MSPE & $R^2_{MSPE}$ & genetic predictors & MSPE & $R^2_{MSPE}$ \\
  \hline
Aggression & Complete case analysis & 601 & 0.083 & 0.261 & 419 & 0.086 & 0.235 \\ 
           & Single imputation & 965 & 0.066 & 0.403 & 744 & 0.069 & 0.376 \\ 
  Hyperactivity & Complete case analysis & 771 & 0.147 & 0.304 & 533 & 0.147 & 0.301 \\ 
                 & Single imputation & 1080 & 0.119 & 0.420 & 905 & 0.121 & 0.411 \\ 
  Opposition & Complete case analysis & 851 & 0.135 & 0.310 & 628 & 0.138 & 0.294 \\ 
             & Single imputation & 1163 & 0.109 & 0.441 & 966 & 0.109 & 0.439 \\ 
   \hline
\end{tabular}
\begin{tablenotes}
      \small
      \item MSPE = Mean squared prediction error.
    \end{tablenotes}
\end{threeparttable}
\end{table}

\clearpage
\section{Discussion}
We proposed a methodology for fitting penalized longitudinal mixed models with more than a single random effect to account for random individual effects not attributable to the genetic similarity between individuals. Our proposed model is based on regularized PQL estimation, which does not require making any assumption about the distribution of the outcome, but only the mean-variance relationship. We studied the performance of the AIREML algorithm when simulating population structure and subjects relatedness for continuous and binary traits. We showed that using PC-AiR to calculate PCs that account for genetic correlations due to distant common ancestry and PC-Relate to estimate kinship due to the sharing of more recent ancestors was effective in controlling the relative bias of variance components estimates under the null model of no genetic association. In addition, we showed that the use of a sparse GRM greatly reduced the computational burden of estimating the variance components and fitting the penalized mixed model, while having little impact on the performance of the lasso penalized mixed model in retrieving important predictors. In simulation studies for both continuous and binary longitudinal traits using real genotype data, we demonstrated that our proposed model achieved better precision (lower FDR) than an univariate association test (GMMAT) and that of a lasso penalized model without any random effect. 

We further showed that omitting to add the top PCs as covariates in the model to adjust for population stratification led to an increase in the relative bias of variance components and in the false discovery rate (FDR) of the lasso penalized mixed model. This is due to PC-Relate only measuring the genetic relatedness due to alleles shared identically by descent (IBD) from recent common ancestors, because genetic relatedness due to more distant ancestry have been adjusted for by previously regressing the genotype values on the top PCs. Using a different approach to estimate the kinship coefficients between individuals, such as the REAP~(\cite{Thornton2012}) and RelateAdmix~(\cite{Moltke2013}) methods may circumvent the need to add the top PCs as covariates in the model to adjust for population structure~(\cite{Chen2020}). However, the advantage of PC-Relate compared to the aforementioned methods is that it does not require model-based estimates of individual ancestry and population-specific allele frequencies nor using external reference population panels. 

Albeit consistent estimation of variance components in mixed models is often overlooked in genetic association studies, we studied the impact of different modelling strategies on the relative bias of the estimates, and showed that increased relative biases were associated with an increase of the FDRs in the penalized model. Since the total phenotypic variance is usually expressed as the sum of polygenic and residual variances, when the polygenic variance is overestimated, the model fails to capture the correct extent of the genetic influences on the phenotype. Furthermore, the variance attributable to the error terms or to within-individual fluctuations in longitudinal studies will be underestimated, leading to potentially inflated type I error rates. In our approach, as we do not test for individual significance of the predictors, this would be translated in observing an increase of the FDR, as was shown in the simulation studies. Additionally, we showed in~\citet{StPierre2023} that the inverse polygenic variance component $\tau^{-1}$ can be seen as a ridge regularization parameter for penalization of the individual polygenic random intercepts. Thus, overestimating the polygenic variance component $\tau$ results in overfitting of the contribution of the PCs obtained from the spectral decomposition of the GRM in explaining the observed phenotypic variance. In other words, it means that the model is overestimating the heterogeneity of the individual random intercepts. From a bias-variance tradeoff point of view, overestimation of the polygenic variance component results in increasing the variability of the random effects estimation, which will be reflected in higher error rates in independent data sets. 

In a real case study for identifying important genetic predictors of aggression, hyperactivity and opposition externalizing behaviors in childhood and adolescents from the QLSCD and QNTS cohorts, we fitted the GMMAT model and our proposed penalized mixed model and showed that the two methods identified different sets of potentially important SNPs. We performed an analysis based on single imputation method to handle missing data due to non-response or loss to follow-up, and compared the results with a complete case analysis. We found two SNPs (rs116541675, rs145226968) that were genome-wide significant (p-value $<$ $5\times10^{-8}$) for the opposition externalizing behaviour in both the single imputation and complete case analyses. For the hyperactivity and aggression behaviours, we found nine mutually exclusive SNPs that were significant in the complete case analysis only. We further demonstrated the utility of our proposed methodology in predicting externalizing behaviours scores in children from the combined QLSCD and QNTS cohorts, and showed using an MSPE-based definitions of the $R^2$ coefficient of determination that the obtained predictions were well-calibrated, and that genetic effects improved the accuracy of the predicted scores compared to a model without any genetic contribution. In addition, we fitted an adaptive lasso penalized mixed model with weights inversely proportional to effect sizes estimates obtained via an elastic-net regularized regression. We found that the the adaptive lasso mixed model resulted in sparser models than the lasso mixed model while the predicted scores accuracy was comparable.

In this study, we have not considered gene-environment interaction (GEI) effects. Conducting genome-wide GEI analyses would contribute to identify
important genetic variants whose effects are modified by environmental factors, and could improve the accuracy of a prediction model as predictors of externalizing behaviours are likely heterogeneous among different environmental exposures, such as socio-economic status. To our knowledge, there exists no GEI univariate association test that allows to account for both genetic similarity and correlation between repeated measurements within an individual for both continuous and binary traits. Further work is needed to assess the computational efficiency and evaluate the estimation and selection accuracy of a penalized hierarchichal variable selection method based on regularized PQL for longitudinal outcomes. An interesting idea to explore is the use of adaptive weights combined with a sparse group lasso penalty in order to perform hierarchichal selection of main genetic and GEI effects in longitudinal studies~(\cite{MendezCivieta2020}).

One limitation of the proposed methodology is that PQL estimation results in biased estimators of both regression coefficients and variance components in GLMMs~(\cite{Jang2009}). It would be interesting to explore if first-order and second-order correction procedures that were proposed in the literature to correct for this bias~(\cite{BRESLOW1995,Lin1996}) would increase the performance of the penalized mixed model in retrieving important predictors when between and within individuals correlations are important. Moreover, it is known that estimated effects by lasso will have large biases because the resulting shrinkage is constant irrespective of the magnitude of the effects. Alternative regularizations like the Smoothly Clipped Absolute Deviation (SCAD)~(\cite{Fan2001}) and Minimax Concave Penalty (MCP)~(\cite{MCP}) could be explored. Another limitation of our proposed method is that we cannot directly analyse imputed SNPs in the dosage format as we rely on the \texttt{SnpArrays} package~(\cite{OpenMendel}) developed in the \texttt{julia} programming language which provides computationally efficient routines for reading and manipulating compressed storage of biallelic SNP data. Finally, when pairwise correlations between SNPs in blocks of LD are important, it is known that the lasso has a tendency to only select one variant among a group of correlated variants. In low-dimensional settings,~\citet{FreijeiroGonzlez2021} showed through a simulation study that the adaptive lasso with weights based on elastic-net regression estimates performed well to retrieve causal predictors in different dependence structures. This is the approach we followed in the real case study. Another direction to explore would be replacing the lasso regularization by an elastic-net penalty in our proposed penalized longitudinal mixed model.

\clearpage
\bibliographystyle{natbib}
\bibliography{biblio}

\clearpage
\begin{appendix}
\setcounter{table}{0}
\setcounter{figure}{0}

\section{Supplementary Tables}\label{appendix:1}
\begin{table}[ht]
\caption{Mean and standard deviation of the relative bias (\%) of variance parameters estimated under the null model of no genetic association when simulating continuous phenotypes with no causal predictor.}
\label{supptab:1}
\centering
\begin{tabular}{llccc}
  \hline
  & & \multicolumn{3}{c}{Number of PCs} \\ \cline{3-5} 
GRM & Variable & \multicolumn{1}{c}{0} & \multicolumn{1}{c}{10} & \multicolumn{1}{c}{20} \\ 
  \hline
full & $\phi$ & 0.13 (1.89) & 0.14 (1.89) & 0.15 (1.89) \\ 
   & $\psi_1$ & -0.49 (12.5) & 0.01 (10.8) & -0.10 (10.6) \\ 
   & $\psi_2$ & 0.89 (12.5) & 1.05 (12.2) & 1.10 (12.2) \\ 
   & $\psi_3$ & -1.49 (20.2) & -1.43 (20.3) & -1.86 (20.3) \\ 
   & $\psi_4$ & 0.55 (8.28) & 0.59 (7.94) & 0.55 (7.97) \\ 
   & $\psi_5$ & -0.20 (9.66) & -0.41 (9.94) & -0.43 (10.1) \\ 
   & $\psi_6$ & 0.97 (6.22) & 1.04 (6.26) & 0.99 (6.30) \\ 
   & $\tau$ & 33.0 (27.2) & 1.64 (13.7) & 1.19 (13.5) \\ \\
  sparse & $\phi$ & 0.12 (1.89) & 0.14 (1.91) & 0.15 (1.90) \\
   & $\psi_1$ & -30.6 (23.6) & -5.86 (20.0) & -4.86 (21.2) \\ 
   & $\psi_2$ & 6.29 (24.2) & 1.16 (12.3) & 1.27 (12.3) \\ 
   & $\psi_3$ & -5.53 (28.2) & -1.28 (20.2) & -1.68 (20.2) \\ 
   & $\psi_4$ & 5.76 (14.2) & 1.01 (9.11) & 0.53 (8.31) \\ 
   & $\psi_5$ & -0.55 (9.89) & -0.68 (10.2) & -0.68 (10.3) \\ 
   & $\psi_6$ & 10.1 (17.3) & 1.84 (8.03) & 1.12 (6.47) \\ 
   & $\tau$ & 72.7 (44.6) & 8.68 (25.2) & 6.35 (26.0) \\ 
   \hline
\end{tabular}
\end{table}

\begin{table}[ht]
\caption{Mean and standard deviation of the relative bias (\%) of variance parameters estimated under the null model of no genetic association when simulating continuous phenotypes with 100 causal predictors explaining 2\% of heritability.}
\label{supptab:2}
\centering
\begin{tabular}{lllll}
   \hline
  & & \multicolumn{3}{c}{Number of PCs} \\ \cline{3-5} 
GRM & Variable & \multicolumn{1}{c}{0} & \multicolumn{1}{c}{10} & \multicolumn{1}{c}{20} \\ 
  \hline
full & $\phi$ & -0.08 (2.04) & -0.09 (1.99) & -0.09 (1.98) \\ 
   & $\psi_1$ & 9.42 (14.4) & 8.99 (13.9) & 9.25 (14.0) \\ 
   & $\psi_2$ & -1.11 (16.0) & -1.85 (14.9) & -1.68 (15.1) \\ 
   & $\psi_3$ & -2.58 (18.6) & -4.16 (17.9) & -4.41 (17.9) \\ 
   & $\psi_4$ & -0.24 (9.24) & 0.14 (8.89) & 0.07 (8.96) \\ 
   & $\psi_5$ & -2.24 (11.1) & -1.79 (10.9) & -1.79 (11.0) \\ 
   & $\psi_6$ & 0.15 (8.17) & 0.17 (7.89) & 0.14 (7.85) \\ 
   & $\tau$ & 46.3 (27.0) & 13.5 (15.8) & 12.6 (15.9) \\ \\
  sparse & $\phi$ & -0.06 (2.01) & -0.08 (1.97) & -0.08 (1.96) \\ 
   & $\psi_1$ & -26.7 (21.3) & -5.35 (25.0) & -4.45 (24.8) \\ 
   & $\psi_2$ & 1.11 (17.8) & -1.59 (15.2) & -1.47 (15.5) \\ 
   & $\psi_3$ & -4.17 (20.6) & -5.26 (19.3) & -5.53 (19.3) \\ 
   & $\psi_4$ & 6.69 (13.8) & 0.54 (9.71) & 0.52 (9.79) \\ 
   & $\psi_5$ & -1.89 (11.4) & -1.57 (11.1) & -1.51 (11.1) \\ 
   & $\psi_6$ & 12.1 (19.4) & 1.02 (9.29) & 1.01 (9.26) \\ 
   & $\tau$ & 95.4 (47.2) & 30.5 (29.6) & 28.90 (29.0) \\ 
   \hline
\end{tabular}
\end{table}

\begin{table}[ht]
\caption{Mean and standard deviation of the relative bias (\%) of variance parameters estimated under the null model of no genetic association when simulating binary phenotypes with no causal predictor.}
\label{supptab:3}
\centering
\begin{tabular}{llccc}
  \hline
  & & \multicolumn{3}{c}{Number of PCs} \\ \cline{3-5} 
GRM & Variable & \multicolumn{1}{c}{0} & \multicolumn{1}{c}{10} & \multicolumn{1}{c}{20} \\ 
  \hline
   full & $\psi_1$ & -7.29 (44.8) & 16.2 (62.4) & 16.2 (63.5) \\ 
   & $\psi_2$ & 44.1 (32.1) & 51.6 (38.5) & 51.4 (38.7) \\ 
   & $\psi_3$ & -18.4 (40.7) & -6.85 (41.1) & -8.69 (40.5) \\ 
   & $\psi_4$ & -9.87 (32.0) & -19.4 (60.7) & -19.8 (60.6) \\ 
   & $\psi_5$ & -26.5 (30.1) & -35.4 (44.8) & -35.6 (44.9) \\ 
   & $\psi_6$ & -42.0 (20.8) & -36.3 (52.7) & -36.4 (51.8) \\ 
   & $\tau$ & 73.7 (47.2) & 41.9 (37.3) & 42.3 (37.6) \\ \\
   sparse & $\psi_1$ & -54.6 (50.7) & 4.18 (57.1) & 7.17 (56.4) \\ 
   & $\psi_2$ & 72.3 (36.2) & 51.8 (37.5) & 48.4 (36.6)  \\ 
   & $\psi_3$ & -57.9 (44.7) & -19.2 (35.3) & -14.3 (37.6) \\ 
   & $\psi_4$ & -12.2 (30.6) & -18.1 (57.3) & -16.3 (56.1) \\ 
   & $\psi_5$ & -23.6 (28.5) & -36.2 (42.1) & -33.8 (41.5) \\ 
   & $\psi_6$ & -39.8 (26.5) & -36.9 (54.5) & -35.3 (54.7) \\ 
   & $\tau$ & 136 (81.7) & 49.9 (57.2) & 46.6 (60.3) \\ 
   \hline
\end{tabular}
\end{table}

\begin{table}[ht]
\caption{Mean and standard deviation of the relative bias (\%) of variance parameters estimated under the null model of no genetic association when simulating binary phenotypes with 100 causal predictors explaining 2\% of heritability.}
\label{supptab:4}
\centering
\begin{tabular}{llccc}
  \hline
  & & \multicolumn{3}{c}{Number of PCs} \\ \cline{3-5} 
GRM & Variable & \multicolumn{1}{c}{0} & \multicolumn{1}{c}{10} & \multicolumn{1}{c}{20} \\ 
  \hline
  full & $\psi_1$ & -5.20 (39.1) & 9.49 (49.3) & 9.55 (48.3) \\ 
   & $\psi_2$ & 39.4 (25.3) & 44.8 (35.0)  & 46.6 (35.2) \\ 
   & $\psi_3$ & -19.4 (43.4) & -17.2 (45.3) & -15.9 (45.9) \\ 
   & $\psi_4$ & -6.20 (19.8) & -19.0 (41.4) & -21.0 (43.2) \\ 
   & $\psi_5$ & -28.4 (25.2) & -31.2 (37.1) & -34.7 (39.2) \\ 
   & $\psi_6$ & -42.3 (18.8) & -43.4 (21.6) & -45.9 (21.5) \\ 
   & $\tau$ & 96.2 (39.0) & 64.6 (37.7) & 63.5 (37.5) \\ \\
  sparse & $\psi_1$ & -54.4 (65.7) & -13.7 (71.7) & -18.3 (66.0) \\ 
   & $\psi_2$ & 74.1 (32.1) & 61.0 (37.2) & 65.0 (36.3)  \\ 
   & $\psi_3$ & -62.5 (47.6) & -36.7 (52.0) & -38.8 (52.5) \\ 
   & $\psi_4$ & -7.81 (39.8) & -19.3 (50.8) & -24.7 (41.2) \\ 
   & $\psi_5$ & -27.6 (26.9) & -33.3 (37.4) & -36.8 (39.5) \\ 
   & $\psi_6$ & -34.5 (53.7) & -40.4 (54.5) & -49.2 (22.1) \\ 
   & $\tau$ & 169 (71.4) & 95.4 (69.2) & 93.1 (67.8) \\ 
   \hline
\end{tabular}
\end{table}

\begin{table}[]
\centering
\caption{Point estimates of the residual variance $\phi$, polygenic variance component $\tau$ and within-individual random effects variances and covariance parameters $\psi_1$, $\psi_2$ and $\psi_3$ estimated under the null model of no genetic association using the AIREML algorithm.}
\label{tab:varcomps}
\begin{threeparttable}
\begin{tabular}{lcccccc}
\hline
  & \multicolumn{2}{c}{Hyperactivity} & \multicolumn{2}{c}{Aggression} & \multicolumn{2}{c}{Opposition} \\ \hline
         & CCA             & SI              & CCA           & SI            & CCA            & SI            \\
$\hat\phi$   & 0.119           & 0.078           & 0.070         & 0.049         & 0.098          & 0.063         \\
$\hat\tau$   & 0.183           & 0.195           & 0.080         & 0.086         & 0.189          & 0.179         \\
$\hat\psi_1$ & 0.308           & 0.505           & 0.350         & 0.365         & 0.237          & 0.413         \\
$\hat\psi_2$ & 0.006           & 0.013           & 0.010         & 0.011         & 0.010          & 0.015     \\
$\hat\psi_3$ & -0.042          & -0.078          & -0.057        & -0.061        & -0.047         & -0.075 \\ \hline
\end{tabular}
\begin{tablenotes}
      \small
      \item CCA = Complete case analysis. SI = Single imputation.
    \end{tablenotes}
\end{threeparttable}
\end{table}

\begin{table}[]
\small
\centering
\caption{Common SNPs selected by the penalized mixed model for all three externalizing scores.}\label{supptab:5}
\begin{threeparttable}
\begin{tabular}{llclcclccc}
  \hline
     & &    &    &    &   &   & \multicolumn{3}{c}{$\hat\beta$} \\
     \cline{8-10} \\
Analysis & SNP & CHR & POS & A1/A2 & MAF & Gene : Consequence & Hyp & Aggr & Opp \\ 
  \hline
CCA & rs4653589 & 1 & 224688598 & G / A & 0.172 & CNIH3 : Intronic & 0.013 & 0.013 & 0.015 \\ 
  & rs12123482 & 1 & 233105677 & A / G & 0.019 & NTPCR : Missense & -0.006 & -0.052 & -0.034 \\ 
  & rs10491702 & 9 & 2227837 & A / C & 0.049 & LOC107987043 : Intronic & 0.019 & 0.001 & 0.005 \\ \\  
  SI & rs1181883 & 1 & 3677933 & C / T & 0.401 & CCDC27 : Missense & 0.005 & $9$x$10^{-7}$ & 0.002 \\ 
  & rs6702929 & 1 & 25406674 & A / C & 0.437 & None & 0.005 & 0.007 & 0.007 \\ 
  & rs4653589 & 1 & 224688598 & G / A & 0.171 & CNIH3 : Intronic & 0.002 & 0.016 & 0.020 \\  
  & rs12994424 & 2 & 26183142 & T / C & 0.101 & KIF3C : Intronic & 0.002 & 0.010 & 0.006 \\ 
  & rs10179260 & 2 & 226521752 & G / A & 0.085 & NYAP2 : Intronic & -0.007 & -0.003 & -0.017 \\ 
  & rs2292997 & 3 & 183724072 & A / G & 0.099 & ABCC5 : Intronic & -0.004 & -0.006 & -0.022 \\ 
  & rs1250109 & 4 & 1227951 & C / T & 0.161 & CTBP1 : Intronic & -0.007 & -0.027 & -0.002 \\ 
  & rs4865047 & 4 & 56821806 & T / C & 0.096 & CEP135 : Intronic & -0.003 & -0.012 & -0.013 \\ 
  & rs1490794 & 5 & 67168343 & A / G & 0.230 & None & 0.008 & 0.006 & 0.004 \\ 
  & rs4292570 & 6 & 82763865 & C / T & 0.200 & LINC02542 : Intronic & -0.103 & -0.008 & -0.032 \\ 
  & rs9372528 & 6 & 119704829 & T / C & 0.069 & None & 0.015 & 0.0009 & 0.016 \\
  & rs547124 & 11 & 120052554 & A / G & 0.141 & LOC124902773 : Intronic & -0.001 & -0.022 & -0.011 \\ 
  & rs12273131 & 11 & 123742913 & T / C & 0.232 & None & -0.005 & $-3$x$10^{-6}$ & -0.013 \\ 
  & rs6571394 & 14 & 21236181 & T / C & 0.458 & LOC107984671 : Intronic & -0.025 & -0.023 & -0.001 \\ 
  & rs4932232 & 15 & 90114309 & G / A & 0.405 & None & -0.008 & -0.043 & -0.028 \\ 
  & rs2951665 & 17 & 32075015 & T / C & 0.485 & ASIC2 : Intronic & 0.003 & 0.002 & 0.008 \\
   \hline
\end{tabular}
\begin{tablenotes}
      \small
      \item CCA = Complete case analysis. SI = Single imputation. MAF = Minor allele frequency.
      \item Hyp = Hyperactivity. Aggr = Aggression. Opp = Opposition.
      \item Note 1: MAF was reported for the effect allele A1.
    \end{tablenotes}
\end{threeparttable}
\end{table}

\begin{table}[]
\smaller
\centering
\caption{Common SNPs selected by the adaptive penalized mixed model for all three externalizing scores.}\label{supptab:6}
\begin{threeparttable}
\begin{tabular}{llclcclccc}
  \hline
     & &    &    &    &   &   & \multicolumn{3}{c}{$\hat\beta$} \\
     \cline{8-10} \\
Analysis & SNP & CHR & POS & A1/A2 & MAF & Gene : Consequence & Hyp & Aggr & Opp \\ 
  \hline
CCA & \textbf{rs4653589} & 1 & 224688598 & G / A & 0.172 & CNIH3 : Intronic & 0.016 & 0.009 & 0.012 \\ 
  & \textbf{rs12123482} & 1 & 233105677 & A / G & 0.019 & NTPCR : Missense & -0.023 & -0.092 & -0.010 \\ \\ 
  SI & \textbf{rs6702929} & 1 & 25406674 & A / C & 0.437 & None & 0.004 & 0.005 & 0.004 \\ 
  & rs921197 & 1 & 30319889 & G / A & 0.473 & None & -0.0006 & -0.002 & -0.006 \\ 
  & \textbf{rs4653589} & 1 & 224688598 & G / A & 0.171 & CNIH3 : Intronic & 0.008 & 0.010 & 0.013 \\ 
  & rs12123482 & 1 & 233105677 & A / G & 0.020 & NTPCR : Missense & -0.019 & -0.034 & -0.037 \\ 
  & rs13027447 & 2 & 139943532 & C / T & 0.181 & None & 0.018 & 0.004 & 0.013 \\ 
  & \textbf{rs10179260} & 2 & 226521752 & G / A & 0.085 & NYAP2 : Intronic & -0.001 & -0.0008 & -0.016 \\ 
  & rs9819889 & 3 & 134584764 & A / G & 0.322 & EPHB1 : Intronic & 0.0001 & 0.018 & 0.012 \\ 
  & rs11922733 & 3 & 183260759 & T / C & 0.057 & KLHL6 : Intronic & 0.003 & 0.0003 & 0.008 \\ 
  & \textbf{rs1250109} & 4 & 1227951 & C / T & 0.161 & CTBP1 : Intronic & -0.011 & -0.028 & -0.002 \\ 
  & rs4835163 & 4 & 150348995 & T / C & 0.143 & IQCM : Intronic & 0.013 & 0.007 & 0.009 \\ 
  & rs7716386 & 5 & 9964861 & A / G & 0.183 & LOC107986405 : Intronic & -0.0001 & -0.036 & -0.028 \\ 
      & \textbf{rs4292570} & 6 & 82763865 & C / T & 0.200 & LINC02542 : Intronic & -0.100 & -0.027 & -0.035 \\ 
   & rs2986977 & 10 & 7950558 & A / G & 0.255 & TAF3 : Intronic & -0.013 & -0.004 & -0.015 \\ 
  & \textbf{rs547124} & 11 & 120052554 & A / G & 0.141 & LOC124902773 : Intronic & -0.004 & -0.023 & -0.009 \\ 
  & rs7138693 & 12 & 75257224 & G / T & 0.278 & LOC105369842 : Non coding & -0.015 & -0.023 & -0.008 \\ 
  & \textbf{rs6571394} & 14 & 21236181 & T / C & 0.458 & LOC107984671 : Intronic & -0.053 & -0.050 & -0.010 \\ 
  & \textbf{rs4932232} & 15 & 90114309 & G / A & 0.405 & None & -0.007 & -0.045 & -0.020 \\ 
   \hline
\end{tabular}
\begin{tablenotes}
      \small
      \item CCA = Complete case analysis. SI = Single imputation. MAF = Minor allele frequency. 
      \item Hyp = Hyperactivity. Aggr = Aggression. Opp = Opposition.
      \item SNPs in bold are SNPs that were also selected by the penalized mixed model (Table \ref{tab:6}).
      \item Note 1: MAF was reported for the effect allele A1.
    \end{tablenotes}
\end{threeparttable}
\end{table}

\clearpage

\section{Supplementary Figures}\label{appendix:2}

\begin{figure}[h]
\centering
\caption{Relative bias of variance parameters estimated under the null model of no genetic association when simulating binary phenotypes with no causal predictor.}
\includegraphics[scale=0.55]{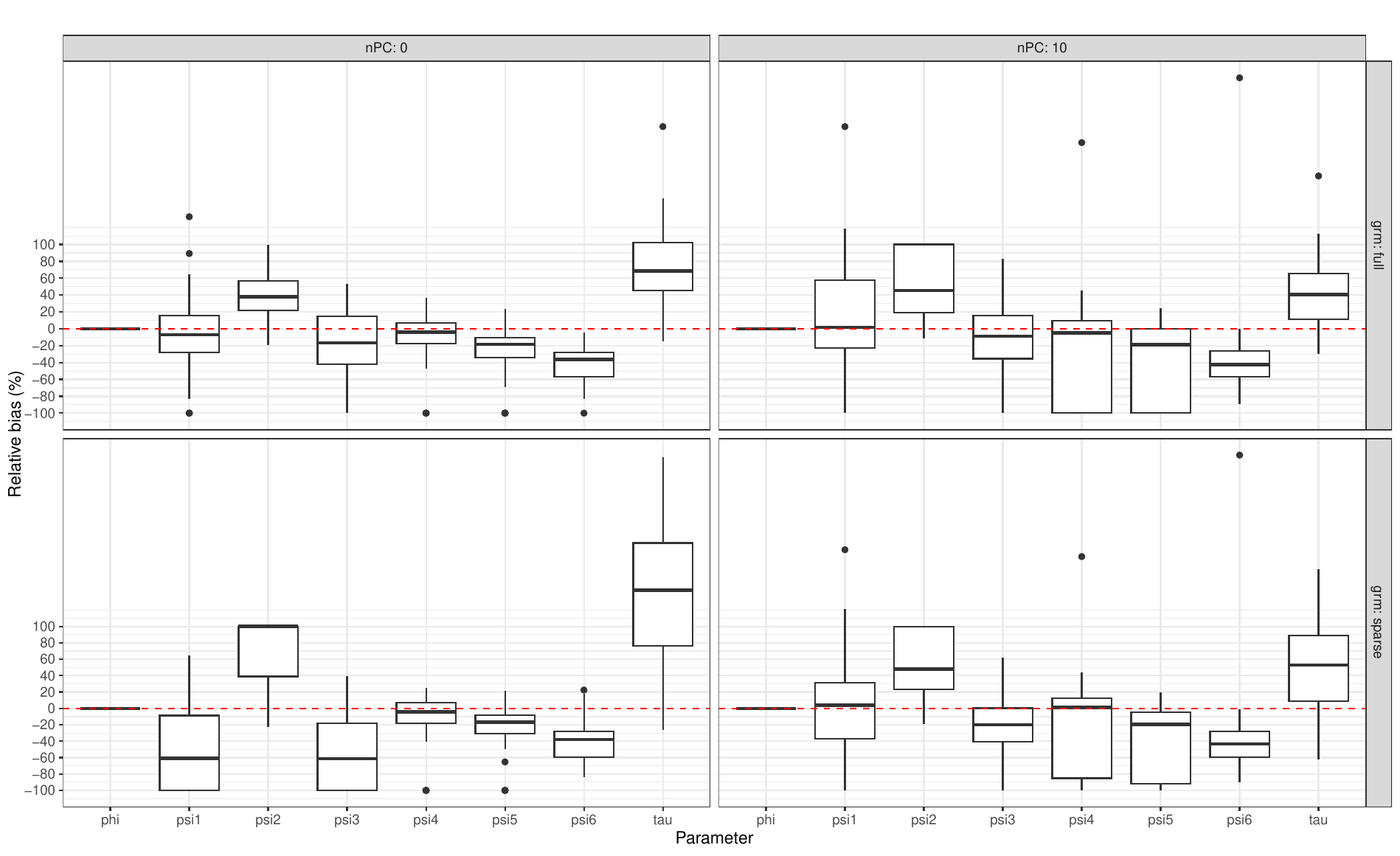}
\label{suppfig:1}
\end{figure}

\begin{figure}[h]
\centering
\caption{Relative bias of variance parameters estimated under the null model of no genetic association when simulating binary phenotypes with 100 causal predictors explaining 2\% of heritability.}
\includegraphics[scale=0.55]{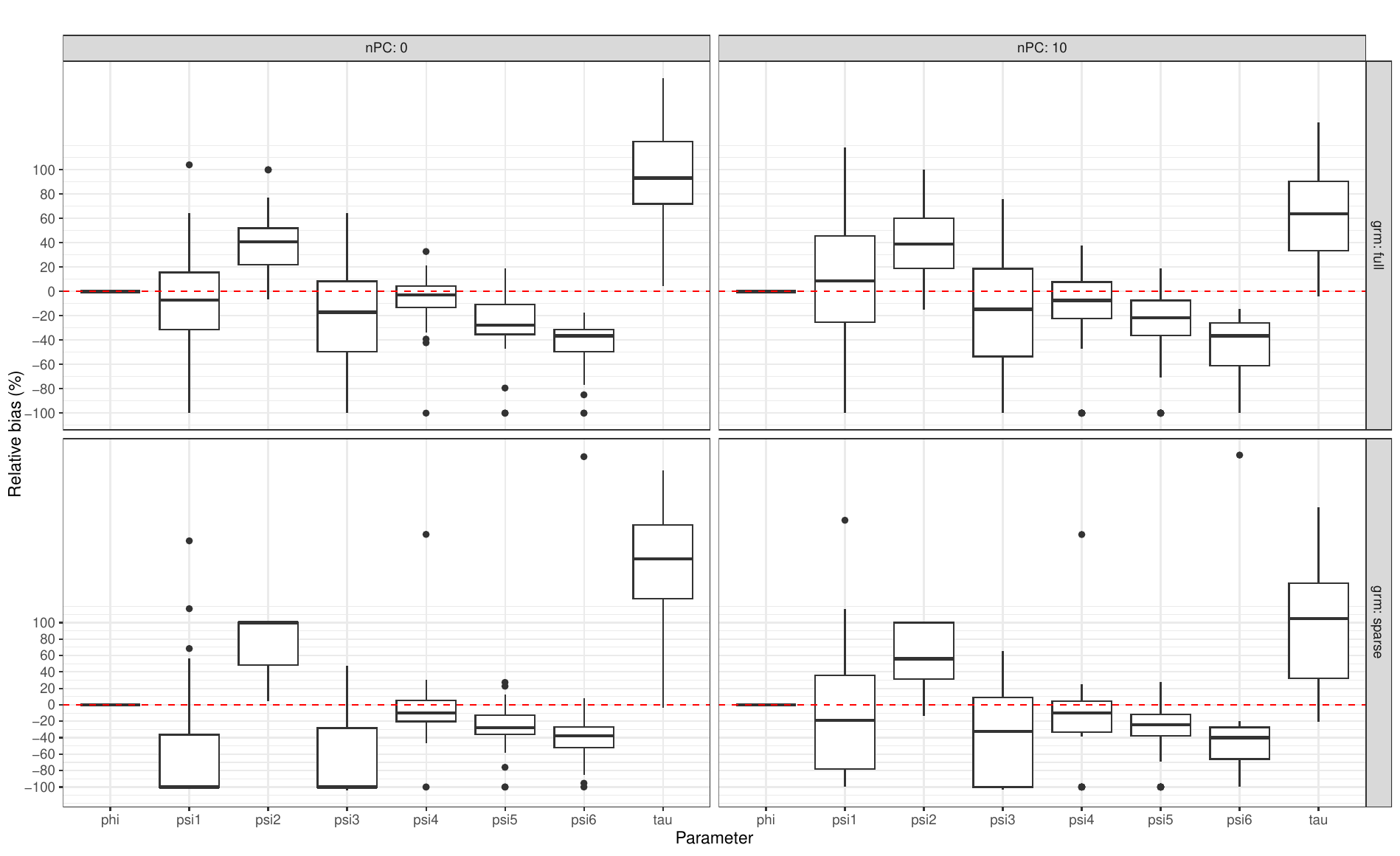}
\label{suppfig:2}
\end{figure}

\begin{figure}[t]
\centering
\caption{Precision-recall curve for selection of genetic predictors for our proposed method as a function of the modelling strategy. The left and right panels illustrate the average performance of the method over 50 replications for the simulation model with binary phenotypes and 100 causal predictors explaining 2\% and 10\% of heritability respectively.}
\includegraphics[scale=0.6]{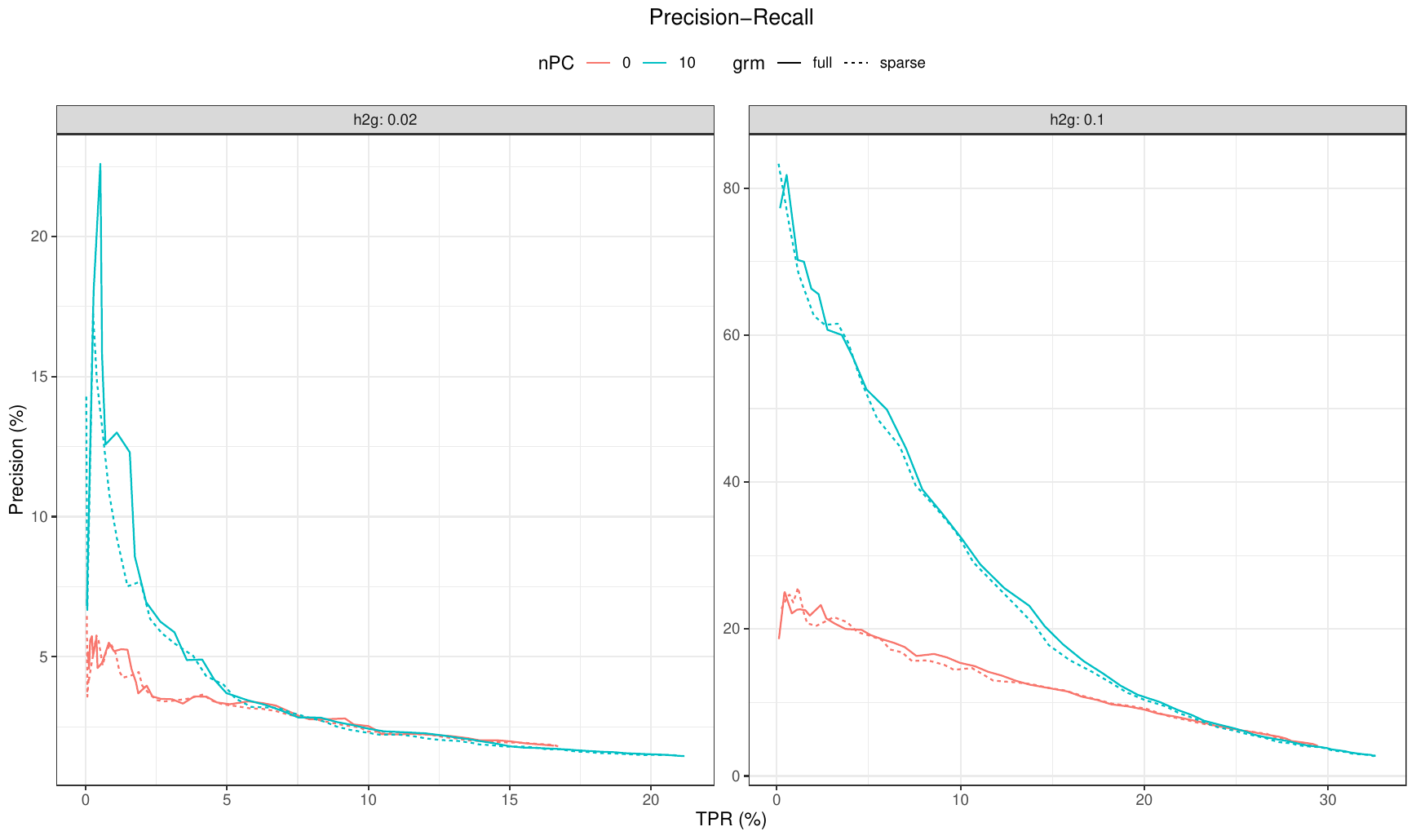}
\label{suppfig:3}
\end{figure}

\begin{figure}[t]
\centering
\caption{Precision-recall curve for selection of genetic predictors for the three compared methods. The left and right panels illustrate the average performance with 95\% confidence interval of the methods over 50 replications for the simulation model with binary phenotypes and 100 causal predictors explaining 2\% and 10\% of heritability respectively.}
\includegraphics[scale=0.6]{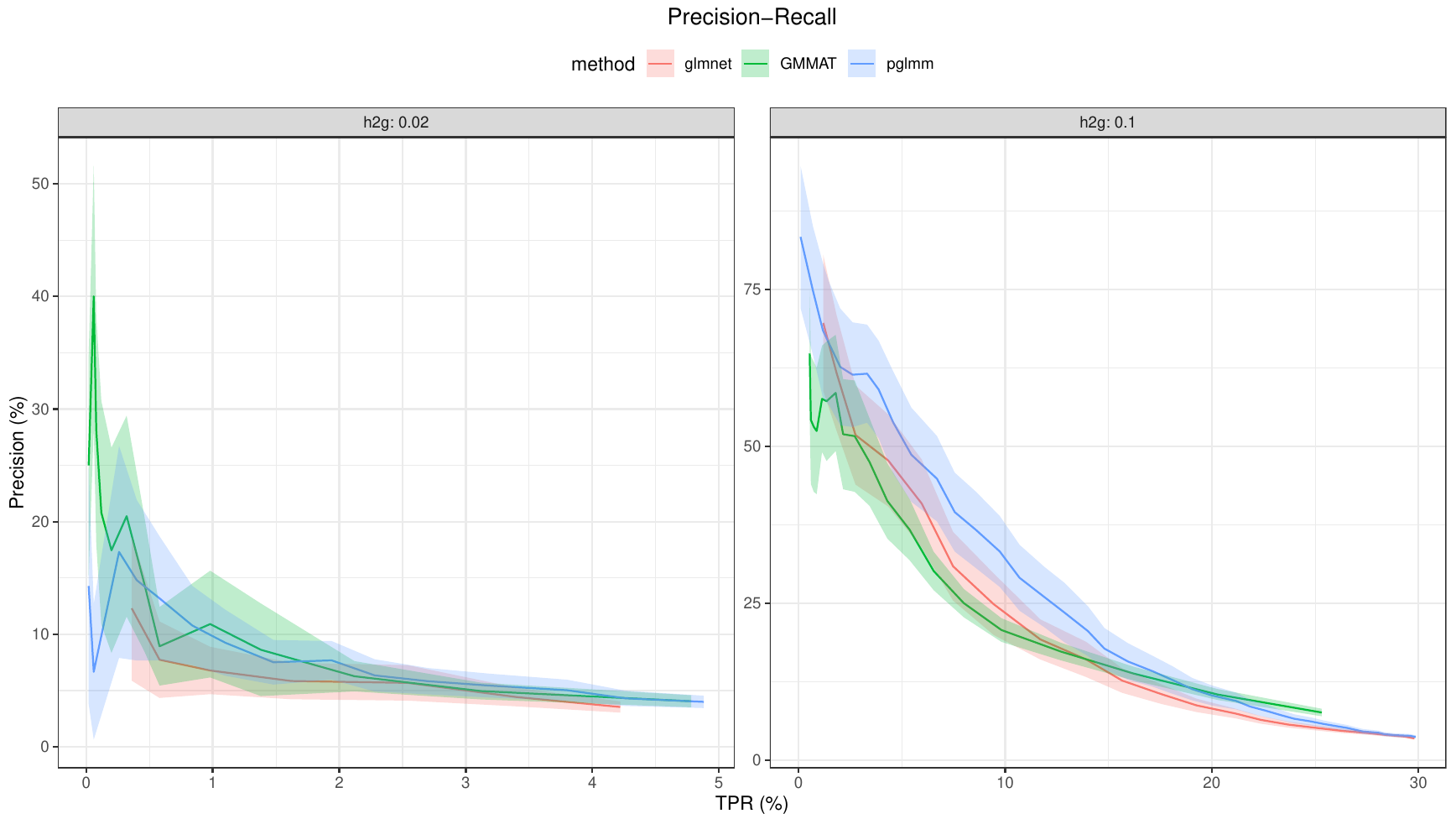}
\label{suppfig:4}
\end{figure}

\begin{figure}[t]
\centering
\caption{Performance of the mixed lasso prediction model ($R^2_{MSPE}$) on the test set as a function of the number of selected genetic predictors. The left and right panels are respectively for the complete case analysis (CCA) and single imputation (SI) model. The dashed vertical lines represent the best model for each externalizing score.}
\includegraphics[scale=0.50]{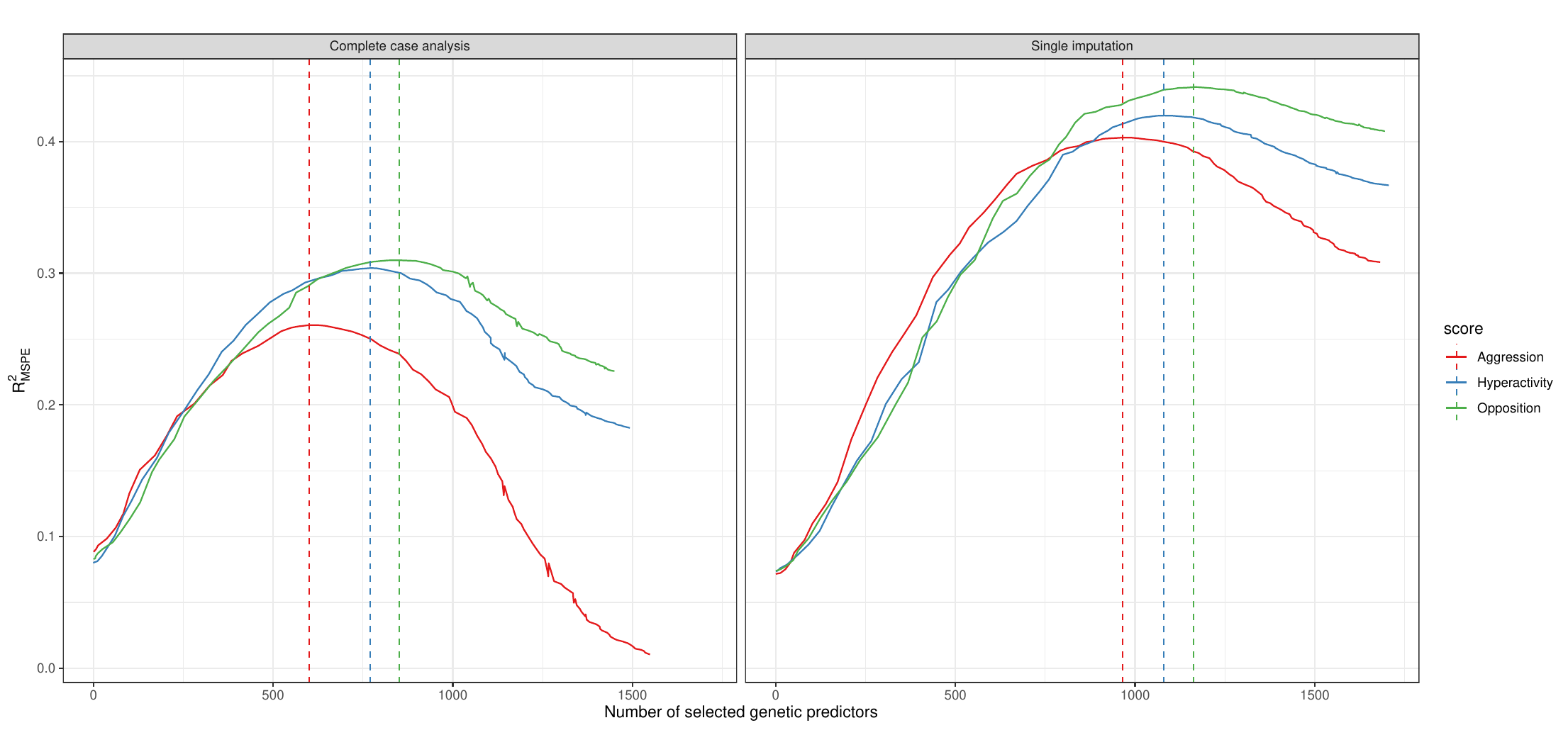}
\label{suppfig:5}
\end{figure}

\begin{figure}[t]
\centering
\caption{Performance of the adaptive mixed lasso prediction model ($R^2_{MSPE}$) on the test set as a function of the number of selected genetic predictors. The left and right panels are respectively for the complete case analysis (CCA) and single imputation (SI) model. The dashed vertical lines represent the best model for each externalizing score. Adaptive weights were estimated by fitting an elastic-net model on the training data. }
\includegraphics[scale=0.50]{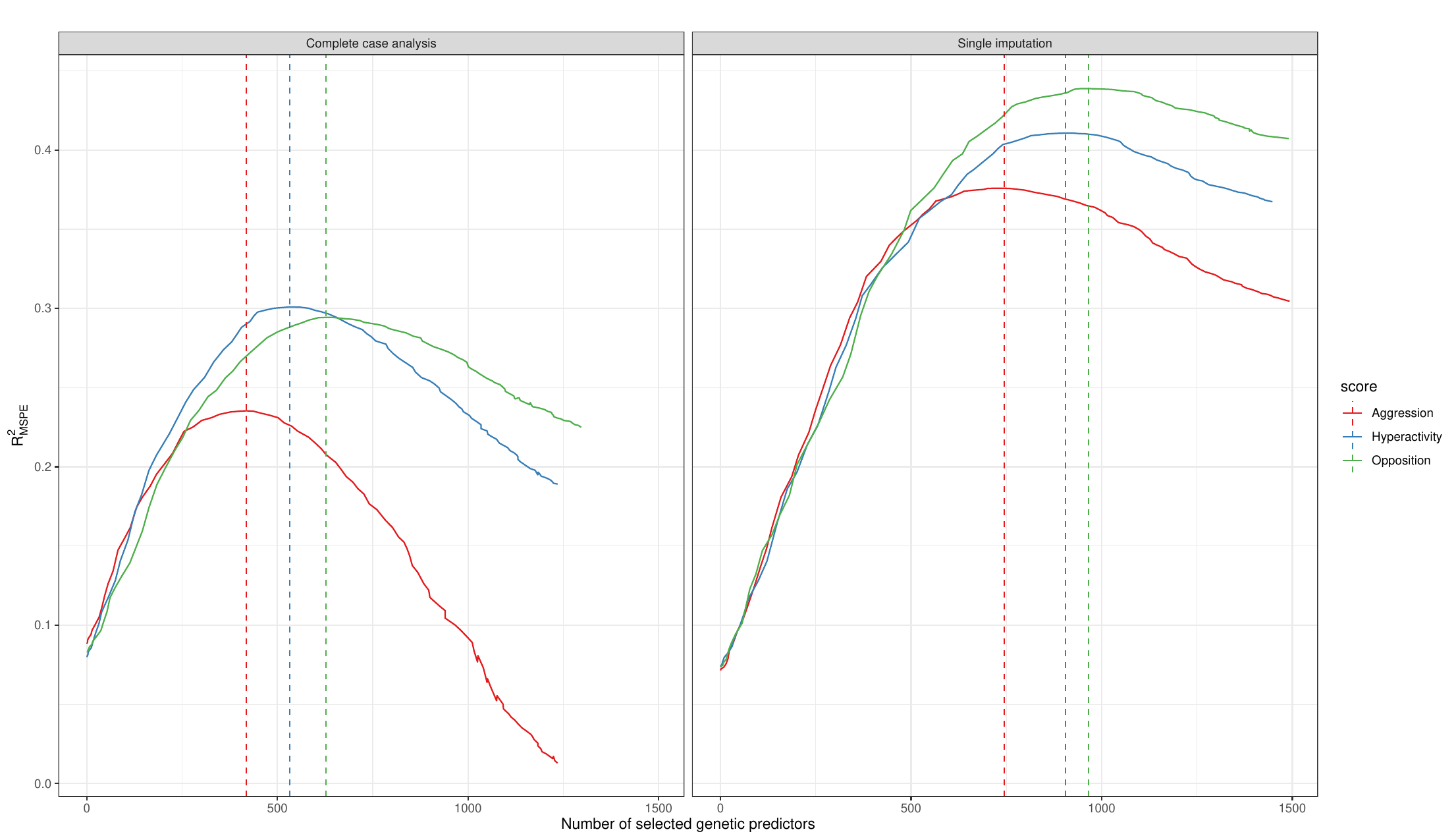}
\label{suppfig:6}
\end{figure}

\clearpage

\section{Genotype quality control}\label{appendix:3}
Genotyping was conducted using the Infinium PsychArray-24 v1.3 BeadChip. The quality control (QC) of genetic data was conducted in PLINK v1.90b5.3, PLINK v1.90b6.7~(\cite{Chang2015}), and R v3.4.3. Pre-imputation QC of genotype data consisted of the following steps:

1. Removal of SNPs with call rates $<98\%$ or a minor allele frequency (MAF) $<1\%$

2. Removal of individuals with genotyping rates $<95\%$ 

3. Removal of sex mismatches

4. Removal of genetic duplicates

5. Removal of cryptic relatives with pi-hat$\ge12.5$

6. Removal of genetic outliers with a distance from the mean of $>4$ SD in the first eight multidimensional scaling (MDS) ancestry components

7. Removal of individuals with a deviation of the autosomal or X-chromosomal heterozygosity from the mean $>4$ SD

8. Removal of non-autosomal variants

9. Removal of SNPs with call rates $<98\%$ or a MAF $<5\%$ or Hardy-Weinberg Equilibrium (HWE) test p-values $<1\times10^{-3}$

10. Removal of A/T and G/C SNPs

11. Update of variant IDs and positions to the IDs and positions in the 1000 Genomes Phase 3 reference panel

12. Alignment of alleles to the reference panel

13. Removal of duplicated variants and variants not present in the reference panel

\end{appendix}

\end{document}


\maketitle

\clearpage

\setcounter{equation}{0}
\renewcommand{\theequation}{A.\arabic{equation}}

\subsection*{Appendix A. PQL Estimation} \label{subsec:pqlest}
The quasi-likelihood for the $jth$ observation from the $ith$ subject given the random effects vector $\mathbf b$ is $$ql_{ij}(\mathbf{\Theta};\mathbf{b}) = \int_{y_{ij}}^{\mu_{ij}}\frac{a_{ij}(y_{ij}-\mu)}{\phi\nu(\mu)} d\mu.$$

The log integrated quasi-likelihood function of $(\mathbf{\Theta}, \phi, \mathbf{\tau}, \mathbf{\psi})$ is 
\begin{align}\label{eq:logintql}
ql(\mathbf{\Theta}, &\phi, \mathbf{\tau}, \mathbf{\psi}) \propto \nonumber \\
& \textrm{log}\int\textrm{exp}\left\{\sum_{i,j}ql_{ij}(\mathbf{\Theta};\mathbf{b}) - \frac{1}{2}  \mathbf{{b}}^\intercal\left(\textrm{diag}\left\{\sum_{k=1}^K \tau_k\mathbf{V}_k , \mathbf D \otimes \mathbf{I}_m\right\}\right)^{-1}\mathbf{b} \right\} \times \left|\sum_{k=1}^K\tau_k\mathbf{V}_k\right|^{-1/2} \left|\mathbf{D}\otimes\mathbf{I}_m\right|^{-1/2} d\mathbf{b}.
\end{align}

Let $$f(\mathbf{b})= \sum_{i,j}ql_{ij}(\mathbf{\Theta};\mathbf{b}) - \frac{1}{2}  \mathbf{{b}}^\intercal\left(\textrm{diag}\left\{\sum_{k=1}^K \tau_k\mathbf{V}_k , \mathbf D \otimes \mathbf{I}_m\right\}\right)^{-1}\mathbf{{b}},$$ we can use Laplace method to approximate the integral $$\int\textrm{exp}\{f(\mathbf{b})\} d\mathbf{b} \approx (2\pi)^{\frac{n+m}{2}}\left|-f''(\tilde{\mathbf{b}})\right|^{-\frac{1}{2}}\textrm{exp}\{f(\tilde{\mathbf{b}})\}, $$ such that equation \eqref{eq:logintql} becomes
\begin{align}\label{eq:ql}
ql(\mathbf{\Theta}, \phi, \mathbf{\tau}, \mathbf{\psi}) = -\frac{1}{2}\textrm{log}\left|\sum_{k=1}^K\tau_k\mathbf{V}_k\right| -\frac{m}{2}\textrm{log}\left|\mathbf{D}\right| -\frac{1}{2}\textrm{log}\left|-f''(\tilde{\mathbf{b}})\right| + f(\tilde{\mathbf{b}}),
\end{align}
where $(\tilde{\mathbf{b}}) = \underset{\mathbf{b}}{\textrm{argmax }}f(\mathbf{b})$ is the solution of $f'(\mathbf{b})=0$.

The first partial derivative of $ql_{ij}(\mathbf{\Theta};\mathbf{b})$ with respect to $\mathbf{b}$ is given by
\begin{align*}
\frac{\partial ql_{ij}}{\partial \mathbf{b}} &= \frac{a_{ij}(y_{ij}-\mu_{ij})}{\phi\nu(\mu_{ij})}\frac{1}{g'(\mu_{ij})}(1, \mathbf{Z}_{ij})^\intercal\otimes\mathbf{L}_{i}^\intercal = \frac{a_{ij}(y_{ij}-\mu_{ij})}{\phi\nu(\mu_{ij})}\frac{1}{g'(\mu_{ij})}\mathbf{H}_{ij}^\intercal,
\end{align*}
where $\mathbf{H}_{ij} = (1, \mathbf{Z}_{ij})\otimes\mathbf{L}_i$ and ${\mathbf{L}}_{i}$ is a $1\times m$ vector of indicators such that $b_{0i}={\mathbf{L}}_{i}\mathbf{b}_0$. For canonical link functions, the second derivative is given by
\begin{align*}
\frac{\partial^2 ql_{ij}}{\partial \mathbf{b}\partial \mathbf{b}^\intercal} &= -\frac{a_{ij}\mathbf{H}_{ij}^\intercal\mathbf{H}_{ij}}{\phi\nu(\mu_{ij})}\frac{1}{g'(\mu_{ij})^2}.
\end{align*}
Let $\mathbf{W} = \phi^{-1}\mathbf{\Delta}^{-1}=\phi^{-1}\textrm{diag}\left\{ \frac{a_{ij}}{\nu(\mu_{ij})[g'(\mu_{ij})^2]}\right\}$ be the diagonal matrix containing weights for each observation, then \eqref{eq:ql} becomes
\begin{align}\label{eq:ql2}
ql(\mathbf{\Theta}, \phi, \mathbf{\tau}, \mathbf{\psi}; \b) &= -\frac{1}{2}\textrm{log}\left|\sum_{k=1}^K\tau_k\mathbf{V}_k\right| -\frac{m}{2}\textrm{log}\left|\mathbf{D}\right| -\frac{1}{2}\textrm{log}\left|\sum_{i,j}\frac{a_{ij}\mathbf{H}_{ij}^\intercal\mathbf{H}_{ij}}{\phi\nu(\mu_{ij})g'(\mu_{ij})^2} + \textrm{diag}\left\{(\sum_{k=1}^K\tau_k\mathbf{V}_k)^{-1}, \mathbf{D}^{-1} \otimes \mathbf{I}_m,\right\}\right| \nonumber \\ 
& \qquad + \sum_{i,j}ql_{ij}(\mathbf{\Theta};\tilde{\mathbf{b}}) - \frac{1}{2}  \tilde{\mathbf{{b}}}^\intercal\left(\textrm{diag}\left\{\sum_{k=1}^K \tau_k\mathbf{V}_k , \mathbf D \otimes \mathbf{I}_m\right\}\right)^{-1}\tilde{\mathbf{{b}}} \nonumber \\
&= -\frac{1}{2}\textrm{log}\left|\textrm{diag}\left\{ \sum_{k=1}^K\tau_k\mathbf{V}_k, \mathbf{D} \otimes \mathbf{I}_m,\right\}\right| -\frac{1}{2}\textrm{log}\left|\mathbf{H}^\intercal\mathbf{WH} + \textrm{diag}\left\{(\sum_{k=1}^K\tau_k\mathbf{V}_k)^{-1}, \mathbf{D}^{-1} \otimes \mathbf{I}_m,\right\}\right| \nonumber \\ 
&\qquad + \sum_{i,j}ql_{ij}(\mathbf{\Theta}; \tilde{\mathbf{b}}) - \frac{1}{2}  \tilde{\mathbf{{b}}}^\intercal\left(\textrm{diag}\left\{\sum_{k=1}^K \tau_k\mathbf{V}_k , \mathbf D \otimes \mathbf{I}_m\right\}\right)^{-1}\tilde{\mathbf{{b}}} \nonumber \\
&= -\frac{1}{2}\textrm{log}\left|\textrm{diag}\left\{\sum_{k=1}^K\tau_k\mathbf{V}_k, \mathbf{D}\otimes\mathbf{I}_m\right\}\mathbf{H}^\intercal\mathbf{WH} + \mathbf{I}_{m(r+1)}\right|
+ \sum_{i,j}ql_{ij}(\mathbf{\Theta}; \tilde{\mathbf{b}}) \nonumber \\ &\qquad - \frac{1}{2}  \tilde{\mathbf{{b}}}^\intercal\left(\textrm{diag}\left\{\sum_{k=1}^K \tau_k\mathbf{V}_k , \mathbf D \otimes \mathbf{I}_m\right\}\right)^{-1}\tilde{\mathbf{{b}}}
\end{align}
where $\mathbf{H}^\intercal = (\mathbf{H}_{11}^\intercal \ \mathbf{H}_{12}^\intercal \ ... \ \mathbf{H}_{mn_m}^\intercal)$ is a $(m(r+1))\times n$ matrix obtained by properly stacking $\mathbf{H}_{ij}$, and $r$ is the dimension of $\mathbf{D}$.

\setcounter{equation}{0}
\renewcommand{\theequation}{B.\arabic{equation}}

\subsection*{Appendix B. Estimation of variance components using AI-REML algorithm}\label{subsec:estcovar}
If $\phi$, $\mathbf\tau$ and $\mathbf \psi$ are known, we jointly choose $\hat{\mathbf{\Theta}}(\phi, \mathbf{\tau}, \mathbf{\psi})$ and $\hat{\mathbf{b}}(\phi, \mathbf{\tau}, \mathbf{\psi})$ to maximize \eqref{eq:ql2}, then $\hat{\mathbf{b}}(\phi, \mathbf{\tau}, \mathbf{\psi}) = \tilde{\mathbf{b}}(\hat{\mathbf{\Theta}}(\phi, \mathbf{\tau}, \mathbf{\psi}))$ because $\tilde{\mathbf{b}}$ maximizes $f(\mathbf{b})$ for given $(\mathbf{\Theta})$. Assuming that the weights in $\mathbf{W}$ vary slowly with the
conditional mean, the derivatives of \eqref{eq:ql2} with respect to $(\mathbf{\Theta}, \mathbf{b})$ are given by
\begin{align*}
\frac{\partial ql(\mathbf{\Theta}, \phi, \mathbf{\tau}, \mathbf{\psi})}{\partial\mathbf{\Theta}} &= \sum_{i=1}^n \frac{a_i(y_i - \mu_i)}{\phi \nu(\mu_i)}\frac{1}{g'(\mu_i)}\mathbf{X}_i^\intercal =\mathbf{X}^\intercal\mathbf{W}\Delta(\mathbf{y}-\mathbf{\mu}), \nonumber \\
\frac{\partial ql(\mathbf{\Theta}, \phi, \mathbf{\tau}, \mathbf{\psi})}{\partial\mathbf{b}} &= 
\sum_{i=1}^n \frac{a_i(y_i - \mu_i)}{\phi \nu(\mu_i)}\frac{1}{g'(\mu_i)}\mathbf{H}_{ij}^\intercal - \left(\textrm{diag}\left\{\sum_{k=1}^K \tau_k\mathbf{V}_k , \mathbf D \otimes \mathbf{I}_m\right\}\right)^{-1}\mathbf{b} \\
&= \mathbf{H}^\intercal\mathbf{W}\Delta(\mathbf{y} - \mathbf{\mu}) - \left(\textrm{diag}\left\{\sum_{k=1}^K \tau_k\mathbf{V}_k , \mathbf{D} \otimes \mathbf{I}_m\right\}\right)^{-1}\mathbf{{b}}, \nonumber
\end{align*}
where $\Delta = \textrm{diag}(g'(\mu_i))$. Defining the working vector $\mathbf{\Y}$ with elements $\tilde{Y_i} = \eta_i + g'(\mu_i)(y_i - \mu_i)$, the solution of 
\begin{gather*}
\begin{cases}
\qquad \mathbf{X}^\intercal\mathbf{W}\Delta(\mathbf{y}-\mathbf{\mu}) =0 \\
\mathbf{H}^\intercal\mathbf{W}\Delta(\mathbf{y} - \mathbf{\mu}) = \left(\textrm{diag}\left\{\sum_{k=1}^K \tau_k\mathbf{V}_k , \mathbf D \otimes \mathbf{I}_m\right\}\right)^{-1}\mathbf{b}
\end{cases}
\end{gather*}
can be written as the solution to the system
\begin{align*}
\begin{bmatrix}
\mathbf{X}^\intercal\mathbf{W}\mathbf{X} & 
\mathbf{X}^\intercal\mathbf{WH} \\
\mathbf{H}^\intercal\mathbf{W}\mathbf{X} & 
\left(\textrm{diag}\left\{\sum_{k=1}^K \tau_k\mathbf{V}_k , \mathbf D \otimes \mathbf{I}_m\right\}\right)^{-1} + \mathbf{H}^\intercal\mathbf{WH}
\end{bmatrix}
\begin{bmatrix}
\mathbf{\Theta} \\ 
\mathbf{b}
\end{bmatrix}
=
\begin{bmatrix}
\mathbf{X}^\intercal\mathbf{W}\mathbf{\Y}\\
\mathbf{H}^\intercal\mathbf{W}\mathbf{\Y}
\end{bmatrix}.
\end{align*}

Let $\mathbf{\Sigma} = \mathbf{H}\textrm{diag}\left\{\sum_{k=1}^K \tau_k\mathbf{V}_k , \mathbf{D} \otimes \mathbf{I}_m\right\} \mathbf{H}^\intercal + \mathbf{W}^{-1} = {\mathbf{L}}(\sum_{k=1}^K\tau_k\mathbf{V}_k){\mathbf{L}}^\intercal + \mathbf{Z}(\mathbf{D}\otimes\mathbf{I}_m)\mathbf{Z}^\intercal + \mathbf{W}^{-1},$
where $\mathbf{Z}^\intercal$ is the $mr\times n$ matrix obtained by properly staking $\mathbf{L}_i^\intercal\otimes\mathbf{Z}_{ij}^\intercal$, and $\mathbf{L}^\intercal$ is the $m\times n$ matrix obtained by properly stacking $\mathbf{L}_i$.
Then, we have
\begin{align*}
&\begin{cases}
\hat{\mathbf{\Theta}} = \left(\mathbf{X}^\intercal\mathbf{\Sigma}^{-1}\mathbf{X}\right)^{-1}\mathbf{X}^{\intercal}\mathbf{\Sigma}^{-1} \mathbf{\Y} \\
\hat{\mathbf{b}}= \textrm{diag}\left\{\sum_{k=1}^K \tau_k\mathbf{V}_k , \mathbf{D} \otimes \mathbf{I}_m\right\}{\mathbf{H}}^\intercal\mathbf{\Sigma}^{-1}\left(\mathbf{\Y}-\mathbf{X}\hat{\mathbf{\Theta}}\right)
\end{cases}.
\end{align*} 
Of note, we have that
\begin{align*}
\mathbf{\Y}-\hat{\mathbf{\eta}}
&=\mathbf{\Y}-\mathbf{X}\hat{\mathbf{\Theta}} - {\mathbf{H}}\hat{\mathbf{b}} \nonumber \\
&= \mathbf{W}^{-1}{\mathbf{\Sigma}}^{-1}\left(\mathbf{\Y}-\mathbf{X}\hat{\mathbf{\Theta}} \right).
\end{align*}

Also, by using the Weinstein–Aronszajn identity, we can show that
\begin{align*}
\left|\textrm{diag}\left\{\sum_{k=1}^K\tau_k\mathbf{V}_k, \mathbf{D}\otimes\mathbf{I}_m\right\}\mathbf{H}^\intercal\mathbf{WH} + \mathbf{I}_{m(r+1)}\right| & =\left|\mathbf{H}\textrm{diag}\left\{ \sum_{k=1}^K\tau_k\mathbf{V}_k, \mathbf{D}\otimes\mathbf{I}_m\right\}\mathbf{H}^\intercal\mathbf{W} + \mathbf{I}_{n}\right| \\
& =\left|({\mathbf{L}}(\sum_{k=1}^K\tau_k\mathbf{V}_k){\mathbf{L}}^\intercal + \mathbf{Z}(\mathbf{D}\otimes\mathbf{I}_m)\mathbf{Z}^\intercal)\mathbf{W} + \mathbf{I}_{n}\right| \\
& =\left|(\mathbf{\Sigma} - \mathbf{W}^{-1})\mathbf{W} + \mathbf{I}_{n}\right| \\
& =\left|\mathbf{\Sigma}\mathbf{W}\right|.
\end{align*}

The log integrated quasi-likelihood function in  \eqref{eq:ql2} evaluated at $(\hat{\mathbf{\Theta}}, \phi, \mathbf{\tau}, \mathbf{\psi})$ becomes
\begin{align*}
ql(\hat{\mathbf{\Theta}},\phi, \mathbf{\tau}, \mathbf{\psi}) =  &-\frac{1}{2}\textrm{log}\left|\mathbf{\Sigma W}\right| -\frac{1}{2}\sum_{i,j}\frac{a_{ij}(y_{ij}-\hat\mu_{ij})^2}{\phi\nu(\hat\mu_{ij})} - \frac{1}{2}\hat{\mathbf{b}}^\intercal\left(\textrm{diag}\left\{\sum_{k=1}^K \tau_k\mathbf{V}_k , \mathbf D \otimes \mathbf{I}_m\right\}\right)^{-1} \hat{\mathbf{b}} \\
=&-\frac{1}{2}\text{log}\left|{\mathbf{\Sigma}}\mathbf{W}\right| -\frac{1}{2}(\mathbf{\Y} - \hat{\mathbf{\eta}})^\intercal\mathbf{W}(\mathbf{\Y} - \hat{\mathbf{\eta}}) \\
& - \frac{1}{2}\left(\mathbf{\Y}-\mathbf{X}\hat{\mathbf{\Theta}}\right)^\intercal{\mathbf{\Sigma}}^{-1}\left( \mathbf{L}\sum_{k=1}^K\tau_k\mathbf{V}_k\mathbf{L}^\intercal + \mathbf{Z}(\mathbf{I}_m\otimes\mathbf{D})\mathbf{Z}^\intercal \right) {\mathbf{\Sigma}}^{-1}\left(\mathbf{\Y}-\mathbf{X}\hat{\mathbf{\Theta}}\right) \\
&=c -\frac{1}{2}\text{log}\left|{\mathbf{\Sigma}}\right| -\frac{1}{2}\left(\mathbf{\Y}-\mathbf{X}\hat{\mathbf{\Theta}}\right)^\intercal{\mathbf{\Sigma}}^{-1}\left(\mathbf{\Y}-\mathbf{X}\hat{\mathbf{\Theta}}\right).
\end{align*}
Similarly, the restricted maximum likelihood (REML) version is~\citep{Harville1977}                                             
\begin{align*}
ql_R(\hat{\mathbf{\Theta}}, \phi, \mathbf{\tau}, \mathbf{\psi})= 
c_R &-\frac{1}{2}\text{log}\left|{\mathbf{\Sigma}}\right| -\frac{1}{2}\text{log}\left|{\mathbf{X}^{*}}^\intercal{\mathbf{\Sigma}}^{-1}\mathbf{X}^*\right| -\frac{1}{2}\left(\mathbf{\Y}-\mathbf{X}\hat{\mathbf{\Theta}}\right)^\intercal{\mathbf{\Sigma}}^{-1}\left(\mathbf{\Y}-\mathbf{X}\hat{\mathbf{\Theta}}\right),
\end{align*}
where $\mathbf{X}^*$ is a full-rank submatrix of $\mathbf{X}$. The first and second derivative of $ql_R(\hat{\mathbf{\Theta}}, \phi, \mathbf{\tau}, \mathbf{\psi})$ with respect to $\vartheta=(\phi, \mathbf{\tau}, \mathbf{\psi})$ are
\begin{align}
\frac{\partial ql_R(\hat{\mathbf{\Theta}}(\phi, \mathbf{\tau}), \phi, \mathbf{\tau}, \mathbf\psi)}{\partial \vartheta_j} &= \frac{1}{2}\left\{\left(\mathbf{\Y}-\mathbf{X}\hat{\mathbf{\Theta}}\right)^\intercal{\mathbf{\Sigma}}^{-1}\frac{\partial\mathbf{\Sigma}}{\partial\vartheta_j}{\mathbf{\Sigma}}^{-1}\left(\mathbf{\Y}-\mathbf{X}\hat{\mathbf{\Theta}}\right) -tr\left(\mathbf{P}\frac{\partial\mathbf{\Sigma}}{\partial\vartheta_j}\right) \right\}, \label{eq:fderiv} \\
\frac{\partial^2 ql_R(\hat{\mathbf{\Theta}}(\phi, \mathbf{\tau}), \phi, \mathbf{\tau}, \mathbf\psi)}{\partial \vartheta_l \partial \vartheta_j} &= \frac{1}{2}\left\{-2\left(\mathbf{\Y}-\mathbf{X}\hat{\mathbf{\Theta}}\right)^\intercal{\mathbf{\Sigma}}^{-1}\frac{\partial\mathbf{\Sigma}}{\partial\vartheta_l}\mathbf{P}\frac{\partial\mathbf{\Sigma}}{\partial\vartheta_j}{\mathbf{\Sigma}}^{-1}\left(\mathbf{\Y}-\mathbf{X}\hat{\mathbf{\Theta}}\right) -tr\left(\mathbf{P}\frac{\partial\mathbf{\Sigma}}{\partial\vartheta_l}\mathbf{P}\frac{\partial\mathbf{\Sigma}}{\partial\vartheta_j}\right)\right\}, \label{eq:sderiv}
\end{align}
where we define the projection matrix $\mathbf{P}=\mathbf{\Sigma}^{-1}-\mathbf{\Sigma}^{-1}\mathbf{X}^*\left({\mathbf{X}^*}^\intercal\mathbf{\Sigma}^{-1}\mathbf{X}^*\right)^{-1}{\mathbf{X}^*}^\intercal\mathbf{\Sigma}^{-1}.$ The elements of the expected information matrix are
\begin{align*}
E \left(-\frac{\partial^2 ql_R(\hat{\mathbf{\Theta}}(\phi, \mathbf{\tau}, \mathbf \psi), \phi, \mathbf{\tau}, \mathbf\psi)}{\partial \vartheta_l \partial  \vartheta_j} \right) = \frac{1}{2}tr\left(\mathbf{P}\frac{\partial\mathbf{\Sigma}}{\partial\vartheta_l}\mathbf{P}\frac{\partial\mathbf{\Sigma}}{\partial\vartheta_j}\right).
\end{align*}
Recall that in the REML iterative process, $\hat{\mathbf{\vartheta}}$ at the (i+1)th iteraton is updated by $\mathbf{\hat\vartheta}^{(i+1)} = \mathbf{\hat\vartheta}^{(i)} + J(\mathbf{\hat\vartheta}^{(i)})^{-1} S(\mathbf{\hat\vartheta}^{(i)})$, where $S(\mathbf{\vartheta})=\frac{\partial ql_{R}(\mathbf{\vartheta})}{\partial\mathbf{\vartheta}}$ and $J(\mathbf{\vartheta})=-\frac{\partial^2 ql_{R}(\mathbf\vartheta{})}{\partial \mathbf{\vartheta}\partial \mathbf{\vartheta}^\intercal}$. In practice, we replace $\mathbf{J}(\mathbf{\vartheta})$ by the average of the observed information $J(\mathbf{\theta})$ and the expected information
\begin{align*}
\mathbf{AI}_{\vartheta_l\vartheta_j} &= \frac{1}{2}\left(\mathbf{\Y}-\mathbf{X}\hat{\mathbf{\Theta}}\right)^\intercal{\mathbf{\Sigma}}^{-1}\frac{\partial\mathbf{\Sigma}}{\partial\vartheta_l}\mathbf{P}\frac{\partial\mathbf{\Sigma}}{\partial\vartheta_j}{\mathbf{\Sigma}}^{-1}\left(\mathbf{\Y}-\mathbf{X}\hat{\mathbf{\Theta}}\right).
\end{align*}

\setcounter{equation}{0}
\renewcommand{\theequation}{C.\arabic{equation}}

\subsection*{Appendix C. Regularized PQL Estimation} \label{subsec:regpqltest}
We define the following objective function $Q_{\lambda}$ which we seek to minimize with respect to $(\mathbf{\Theta}, \phi, \mathbf{\tau}, \mathbf{\psi}; \b)$:
\begin{align}\label{eq:objfunc}
Q_{\lambda}(\mathbf{\Theta}, \phi, \mathbf{\tau}, \mathbf{\psi}; \b) := -ql(\mathbf{\Theta}, \phi, \mathbf{\tau}, \mathbf{\psi}; \b) + \lambda\sum_j|\beta_j|,
\end{align}
where $\lambda>0$ controls the strength of the overall regularization. 

If $\phi$, $\mathbf\tau$ and $\mathbf \psi$ are known, we solve for $\hat{\mathbf{\Theta}}(\phi, \mathbf{\tau}, \mathbf{\psi})$ and $\hat{\mathbf{b}}(\phi, \mathbf{\tau}, \mathbf{\psi})$ by iteratively minimizing  \eqref{eq:objfunc} with respect to each parameter. Assuming that the weights in $\mathbf{W}$ vary slowly with the
conditional mean, the derivative of \eqref{eq:ql2} with respect to and $\b$ is given by
\begin{align*}
\frac{\partial ql(\mathbf{\Theta}, \phi, \mathbf{\tau}, \mathbf{\psi})}{\partial\b} &= 
\sum_{i=1}^m\sum_{j=1}^{n_i} \frac{a_{ij}(y_{ij} - \mu_{ij})}{\phi \nu(\mu_{ij})}\frac{1}{g'(\mu_{ij})}{\mathbf{H}}_{ij}^\intercal - \left(\textrm{diag}\left\{\sum_{k=1}^K \tau_k\mathbf{V}_k , \mathbf D \otimes \mathbf{I}_m\right\}\right)^{-1}\b
\\
&= {\mathbf{H}}^\intercal\mathbf{W}\Delta(\mathbf{y} - \mathbf{\mu}) - \left(\textrm{diag}\left\{\sum_{k=1}^K \tau_k\mathbf{V}_k , \mathbf D \otimes \mathbf{I}_m\right\}\right)^{-1}\b.
\end{align*}

The second derivative is given by
\begin{align*}
\frac{\partial^2 ql(\mathbf{\Theta}, \phi, \mathbf{\tau}, \mathbf{\psi})}{\partial\b\partial\b^\intercal} &= -\mathbf{H}^\intercal\mathbf{WH} - \left(\textrm{diag}\left\{\sum_{k=1}^K \tau_k\mathbf{V}_k , \mathbf D \otimes \mathbf{I}_m\right\}\right)^{-1}.
\end{align*}

This leads to the Newton updates
\begin{align}\label{eq:Newton1}
\hat{\mathbf{b}}^{(t+1)} &= \hat{\mathbf{b}}^{(t)} - \left(\frac{\partial^2 ql(\mathbf{\Theta}, \phi, \mathbf{\tau}, \mathbf{\psi})}{\partial\b\partial\b^\intercal}\right)^{-1} \frac{\partial ql(\mathbf{\Theta}, \phi, \mathbf{\tau}, \mathbf{\psi})}{\partial\b} \nonumber \\
&= \left(\mathbf{H}^\intercal\mathbf{WH} + \textrm{diag}\left\{(\sum_{k=1}^K \tau_k\mathbf{V}_k)^{-1} , \mathbf D^{-1} \otimes \mathbf{I}_m\right\}\right)^{-1}\mathbf{H}^\intercal\mathbf{W}\left(\mathbf{\Delta}(\mathbf y - \mathbf\mu) + \mathbf{H}\hat{\mathbf{b}}^{(t)}\right).
\end{align}

Defining the working vector $\tilde{\mathbf{Y}} = \mathbf{X}\mathbf{\Theta} + \mathbf{H}{\mathbf{b}} + \mathbf{\Delta}(\mathbf y - \mathbf\mu)$, where $\mathbf{X}\mathbf{\Theta}=\mathbf{C} \mathbf\theta + \mathbf E \alpha + \mathbf{G} \mathbf\beta + (\mathbf D \odot \mathbf{G}) \mathbf{\gamma}$, equation \eqref{eq:Newton1} can be rewritten as
\begin{align}\label{eq:irwls_b}
\hat{\mathbf{b}}^{(t+1)} &=  \underset{\mathbf b}{\textrm{argmin }} \left(\tilde{\mathbf{Y}} - \mathbf{X}\hat{\mathbf{\Theta}} -\mathbf{Hb}\right)^\intercal \mathbf{W} \left(\tilde{\mathbf{Y}} - \mathbf{X}\hat{\mathbf{\Theta}} - \mathbf{Hb} \right) + \mathbf{b}^\intercal\textrm{diag}\left\{(\sum_{k=1}^K \tau_k\mathbf{V}_k)^{-1} , \mathbf D^{-1} \otimes \mathbf{I}_m\right\}\mathbf b \nonumber \\
&=  \underset{\mathbf b}{\textrm{argmin }} \left(\tilde{\mathbf{Y}} - \mathbf{X}\hat{\mathbf{\Theta}} - \mathbf{Hb}\right)^\intercal \mathbf{W} \left(\tilde{\mathbf{Y}} - \mathbf{X}\hat{\mathbf{\Theta}} - \mathbf{Hb} \right) + (\mathbf{U}^\intercal\mathbf b)^\intercal\mathbf{\Lambda}^{-1}\mathbf{U}^\intercal\mathbf b,
\end{align}
where $\mathbf{U}$ is an orthonormal matrix of eigenvectors and $\mathbf \Lambda$ is a diagonal matrix of eigenvalues $\Lambda_1\ge \Lambda_2 \ge...\ge \Lambda_{m(r+1)} > 0$ if $\sum\tau_k\mathbf{V}_k$ and $\mathbf D$ are positive definite. Letting $\mathbf{\delta}=\mathbf{U}^\intercal\mathbf{b}$ and $\mathbf{U_H}=\mathbf{HU}$, minimizing \eqref{eq:irwls_b} is equivalent to minimizing
\begin{align}\label{eq:irwls_delta}
\hat{\mathbf{\delta}}^{(t+1)}
&=  \underset{\mathbf \delta}{\textrm{argmin }} \left(\tilde{\mathbf{Y}} - \mathbf{X}\hat{\mathbf{\Theta}} -\mathbf{U_H}\mathbf{\delta}\right)^\intercal \mathbf{W} \left(\tilde{\mathbf{Y}} - \mathbf{X}\hat{\mathbf{\Theta}} - \mathbf{U_H}\mathbf{\delta} \right) + \mathbf{\delta}^\intercal\mathbf{\Lambda}^{-1}\mathbf{\delta}.
\end{align}

Finally, we use a block coordinate descent algorithm to find PQL regularized estimates for $\mathbf{\Theta}$. Let $\mathbf{{\Theta}}^{(t)}$ be the current iterate, we again rewrite the minimization problem as a weighted least squares (WLS) problem where
\begin{align}\label{eq:proxnewton}
\mathbf{{\Theta}}^{(t+1)}
&= \underset{\mathbf{{\Theta}}}{\textrm{argmin }} \left\{\frac{1}{2s_t} \sum_{i=1}^m\sum_{j=1}^{n_i} \bar{w}_{ij}\left(\tilde{Y}_{ij} - \mathbf{X}_{ij}\mathbf{\Theta} -{\mathbf{U_H}}_{ij}\hat{\mathbf{\delta}}^{(t+1)}\right)^2 + \lambda\sum_j|\beta_j|\right\},
\end{align}
where $\bar{w}_{ij} = \mathbf{\Delta}_{ij}^{-(t)}$. We use block coordinate descent and minimize \eqref{eq:proxnewton} with respect to each component of $\mathbf{\Theta}=\left(\mathbf{\theta}^\intercal, \mathbf\beta^\intercal ,\mathbf{\gamma}^\intercal\right)^\intercal$. In practice, we set $s_t=1$ and do not perform step-size optimization.

\clearpage
\bibliographystyle{apalike}
\bibliography{biblio}